\documentclass[11pt]{article}
\usepackage[final]{acl}
\usepackage{times}
\usepackage{latexsym}
\usepackage[T1]{fontenc}
\usepackage[utf8]{inputenc}
\usepackage{microtype}
\usepackage{inconsolata}
\usepackage{graphicx}
\usepackage{booktabs}
\usepackage{amsmath}
\usepackage{amssymb}
\usepackage{array}
\usepackage{multirow}
\usepackage{xcolor}
\usepackage{enumitem}
\newcommand{\symbolimg}[2][0.26cm]{%
  \ensuremath{\vcenter{\hbox{\includegraphics[height=#1]{#2}}}}}
\newcommand{\ivnova}{\symbolimg[0.25cm]{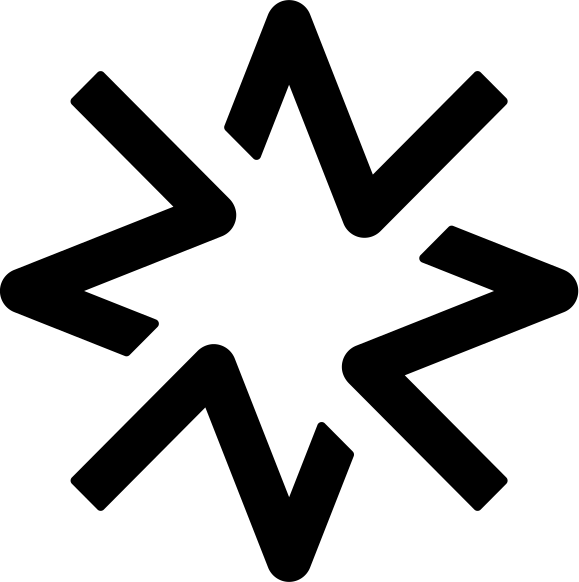}}
\newcommand{\ivgemini}{\symbolimg[0.25cm]{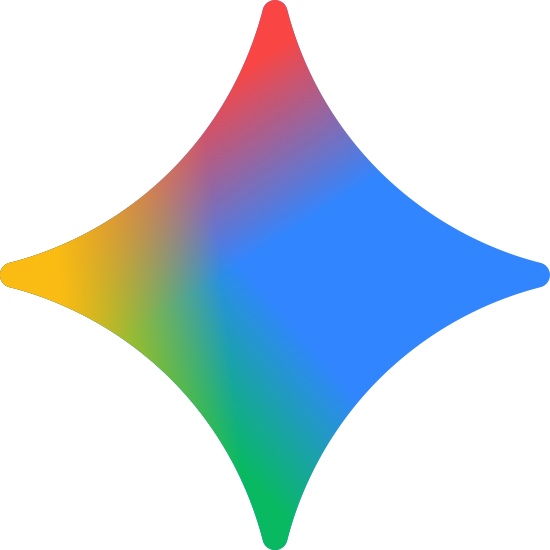}}
\newcommand{\ivopenai}{\symbolimg[0.24cm]{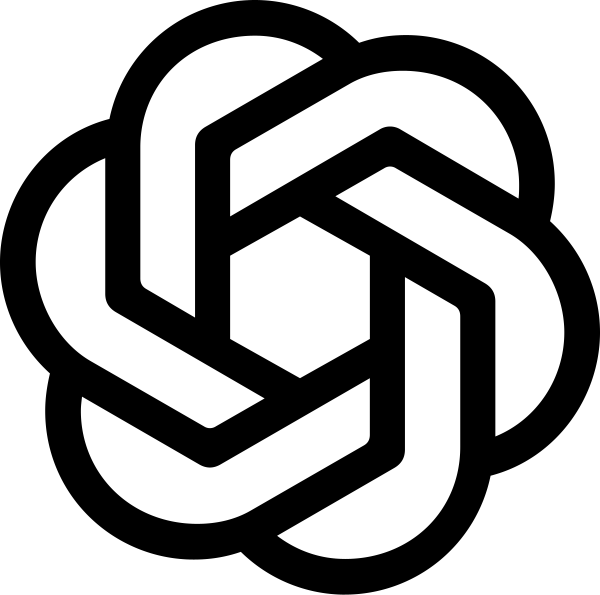}}
\newcommand{\ivgrok}{\symbolimg[0.28cm]{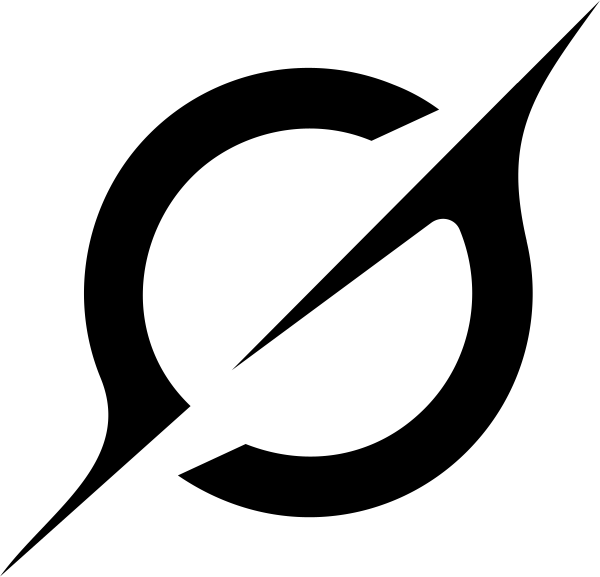}}
\newcommand{\Nova}{\ivnova\,Nova 2 Sonic}
\newcommand{\Gem}{\ivgemini\,Gemini-3.1-Flash-Live}
\newcommand{\Gpt}{\ivopenai\,GPT-Realtime-2.1}
\newcommand{\Gptm}{\ivopenai\,GPT-Realtime-2.1-mini}
\newcommand{\Grok}{\ivgrok\,Grok Voice Think Fast 1.0}
\newcommand{\NovaS}{\ivnova\,2 Sonic}
\newcommand{\GemS}{\ivgemini\,3.1-Flash-Live}
\newcommand{\GptS}{\ivopenai\,Realtime-2.1}
\newcommand{\GptmS}{\ivopenai\,Realtime-2.1-mini}
\newcommand{\GrokS}{\ivgrok\,Voice Think Fast}
\newcommand{\gbank}{\symbolimg[0.24cm]{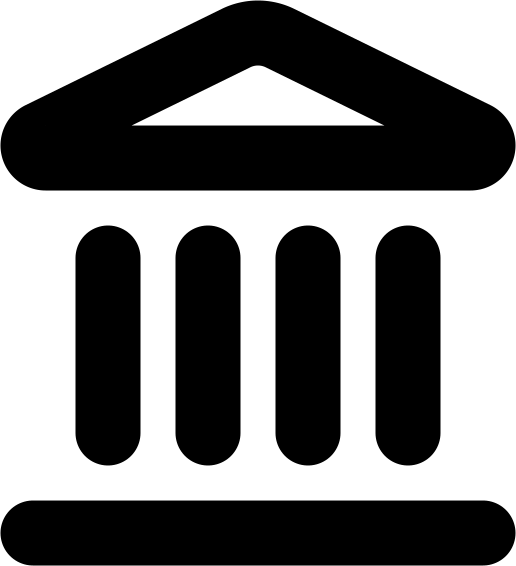}}
\newcommand{\gins}{\symbolimg[0.24cm]{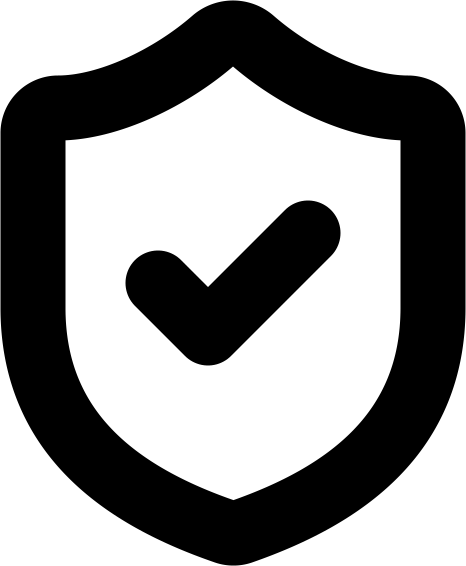}}
\newcommand{\gtrav}{\symbolimg[0.24cm]{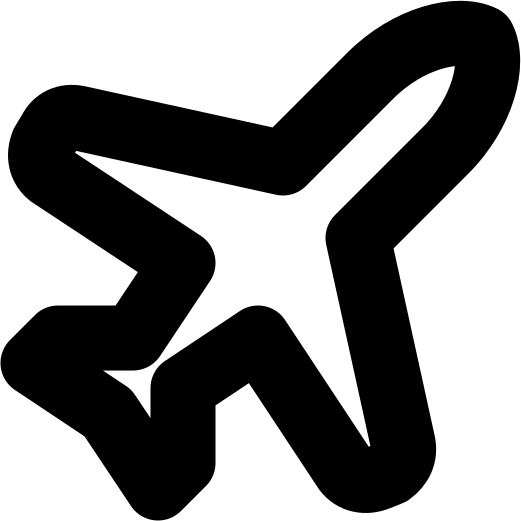}}
\newcommand{\ghealth}{\symbolimg[0.24cm]{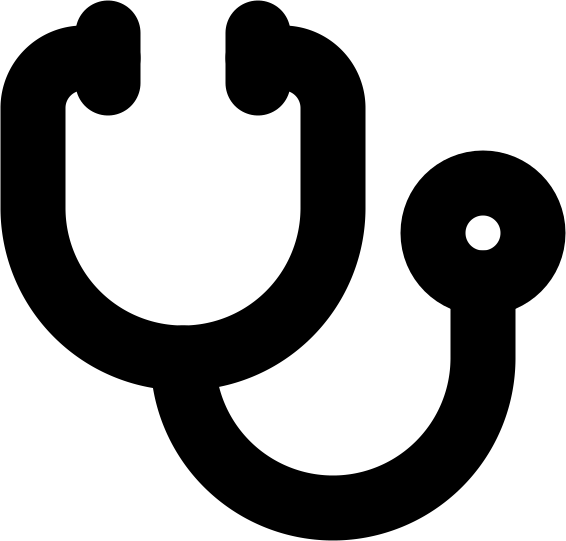}}
\newcommand{\glog}{\symbolimg[0.24cm]{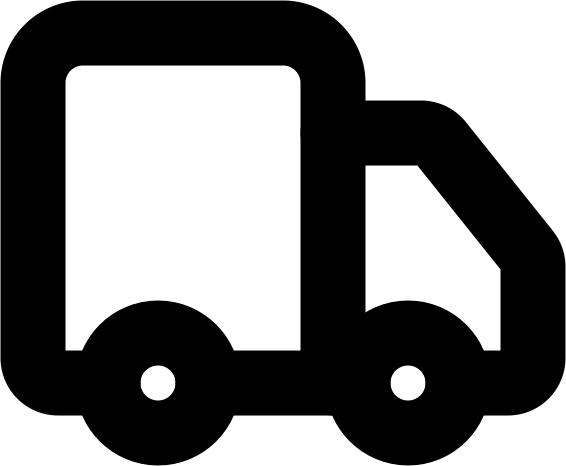}}
\newcommand{\gway}{\symbolimg[0.24cm]{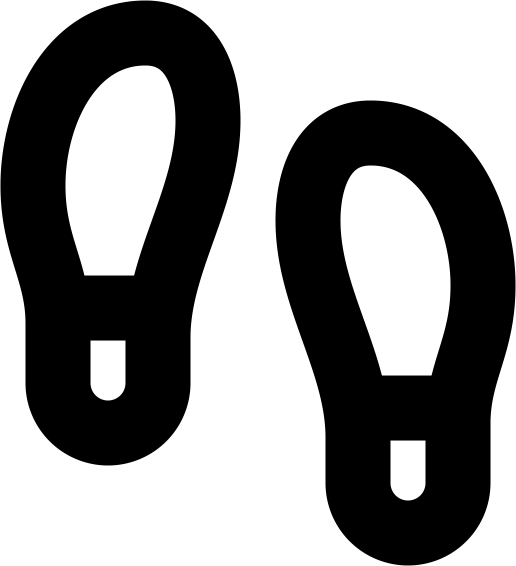}}
\newcommand{\Wbank}{\gbank\,Banking}
\newcommand{\Wins}{\gins\,Insurance}
\newcommand{\Wtrav}{\gtrav\,Travel}
\newcommand{\Whealth}{\ghealth\,Healthcare}
\newcommand{\Wlog}{\glog\,Logistics}
\newcommand{\Wway}{\gway\,Pathfinding}
\newcommand{\gdyn}{\symbolimg[0.24cm]{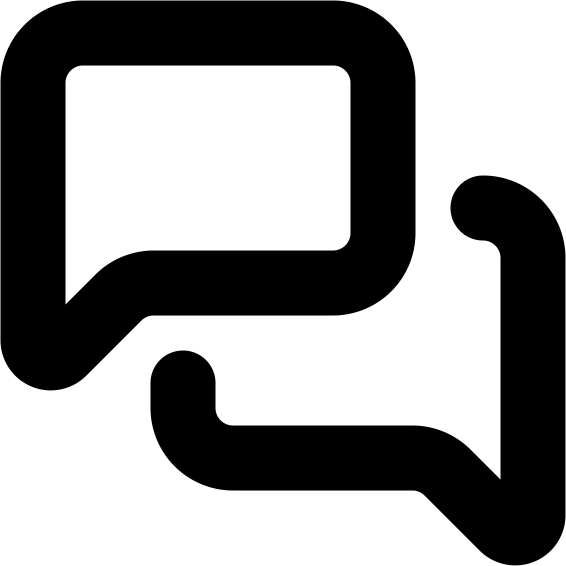}}
\newcommand{\gtool}{\symbolimg[0.24cm]{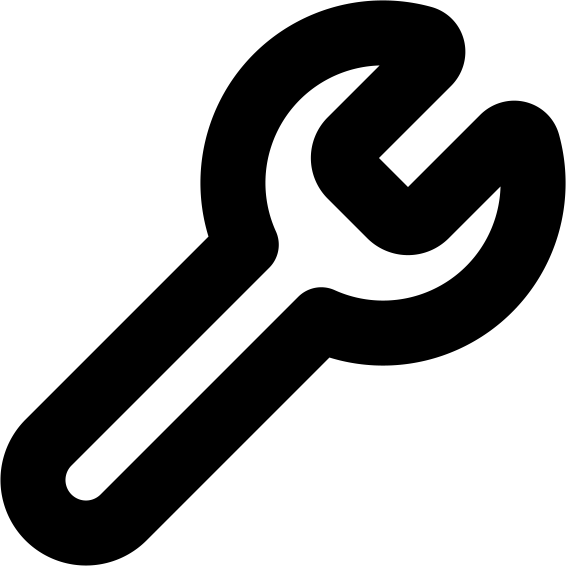}}
\newcommand{\gnat}{\symbolimg[0.24cm]{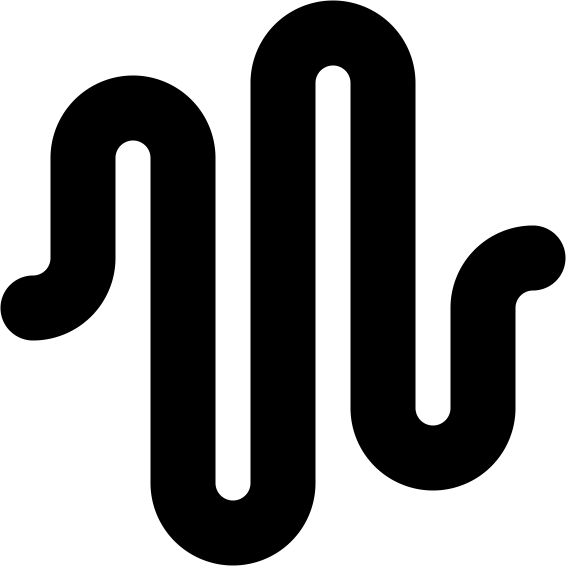}}
\newcommand{\axi}[1]{\symbolimg[0.23cm]{icon-a#1.png}}
\newcommand{\axerrand}{\symbolimg[0.23cm]{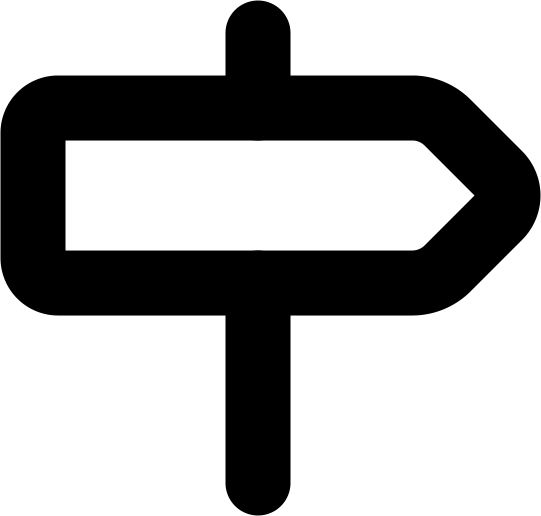}}
\newcommand{\axblocked}{\symbolimg[0.23cm]{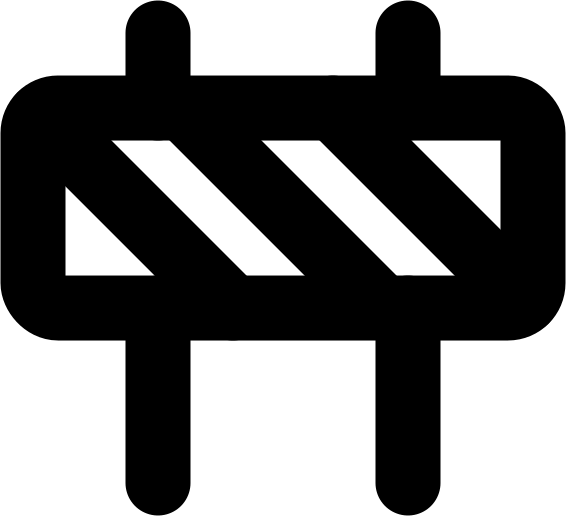}}
\newcommand{\gsym}{\ensuremath{\mathrm{GS}}}
\newcommand{\pat}{\ensuremath{\mathrm{Pass}@3}}
\newcommand{\sNova}{\ivnova$_{\text{2S}}$}
\newcommand{\sGem}{\ivgemini$_{\text{3.1}}$}
\newcommand{\sGpt}{\ivopenai$_{\text{2.1}}$}
\newcommand{\sGptm}{\ivopenai$_{\text{mini}}$}
\newcommand{\sGrok}{\ivgrok$_{\text{VTF}}$}
\newcommand{\bench}{\textsc{DuplexWorld}}

\newcommand{\nworlds}{six}
\newcommand{\domainlist}{banking, insurance, travel, healthcare and logistics}
\newcommand{\navworld}{Pathfinding}
\newcommand{\ntypes}{eleven}

\newcommand{\nscen}{156}
\newcommand{\nscamp}{144}

\newcommand{\nconvnav}{450}
\newcommand{\nconvall}{3{,}825}
\newcommand{\cmark}{\ensuremath{\checkmark}}
\newcommand{\xmark}{\ensuremath{-}}
\newcommand{\pk}{\ensuremath{\mathrm{Pass}^{k}}}
\newcommand{\pone}{\ensuremath{\mathrm{Pass}@1}}
\newcommand{\patk}{\ensuremath{\mathrm{Pass}@k}}
\newcommand{\pthree}{\ensuremath{\mathrm{Pass}^{3}}}
\newcommand{\ci}[1]{\ensuremath{{}_{\pm#1}}}
\title{~\raisebox{-18pt}{\includegraphics[height=40pt]{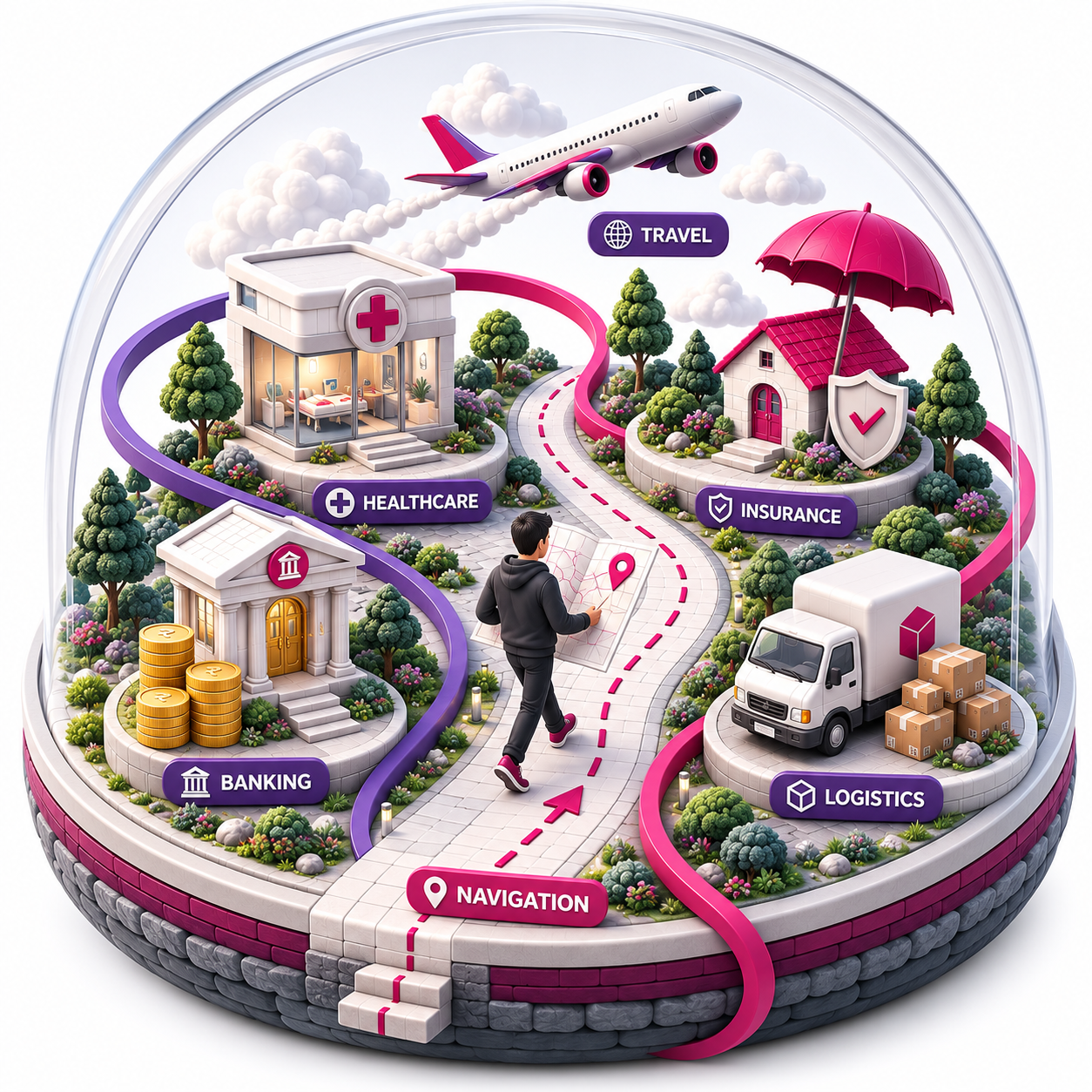}}~\bench{}: Can voice agents help you get through the day?}

\author{
  Aryan Vijay Bhosale$^{1,2\ast}$, Harshit Rajgarhia$^{1\ast}$,
  Akhil Pothanapalli$^{1\ast}$, \\
  \textbf{Asif Shaik}$^{1}$,
  \textbf{Abhishek Mukherji}$^{1\dagger}$,
  \textbf{Dinesh Manocha}$^{2\dagger}$ \\
  $^{1}$\href{https://www.centific.com/research/applied-ai-research}{Centific Global Solutions Inc.} \quad
  $^{2}$\href{https://www.cs.umd.edu/}{University of Maryland} \\
  \texttt{\{aryan.bhosale, harshit.rajgarhia\}@centific.com} \\
  \large\normalfont\url{https://duplexworld.github.io}
}

\begin{document}
\maketitle
{\renewcommand{\thefootnote}{}%
\footnotetext{\hspace*{-1.8em} Work done during an internship at
  \href{https://www.centific.com/research/applied-ai-research}{Centific}. \\ $^{\ast}$Equal contribution.
  $^{\dagger}$Equal advising. 
  }%
\addtocounter{footnote}{-1}}
% {\renewcommand{\thefootnote}{}%
% \footnotetext{$^{\ast}$Equal contribution. $^{\dagger}$Equal advising.}%
% \footnotetext{Work done during an internship at \href{https://www.centific.com/research/applied-ai-research}{Centific}.}}
\begin{abstract}
Speech-to-speech (S2S) voice agents are increasingly being incorporated into enterprise for customer care and as daily companions for consumers owing to the ease of the conversational modality over text. However, existing benchmarks fail to holistically evaluate voice agents along axes that really matter and are shaped as tests of agentic tool calling against a database. We believe they fail to adequately account for the diversity of conversational dialogue that mundane activities introduce and further, never test how faithfully an agent can assist
on tasks that move beyond database manipulation. To tackle this \bench{} introduces \nworlds{} worlds where voice agents are especially useful: \domainlist{}, and \navworld{}. Agents are evaluated on \ntypes{} different types of conversations across \nscen{} scenarios (350$+$ hours of conversation), each testing conversational and analytical capability to varying degrees. Through extensive evaluation comprising agentic, conversational and speech-naturalness metrics, we show that even the best voice agents leave substantial room for improvement on all 3 axes (\pone{}: 0.490, turn-taking: 0.653, DNSMOS: 3.378). We perform extensive analysis on agentic v conversational performance, world- and conversation type-wise performance, failure modes exploring the explore v exploit lens for \navworld{} conversations and voice agent reliability over all \nworlds{} worlds.
\end{abstract}
\section{Introduction}
\label{sec:intro}
\begin{figure}[t]
\centering
\includegraphics[width=0.75\columnwidth]{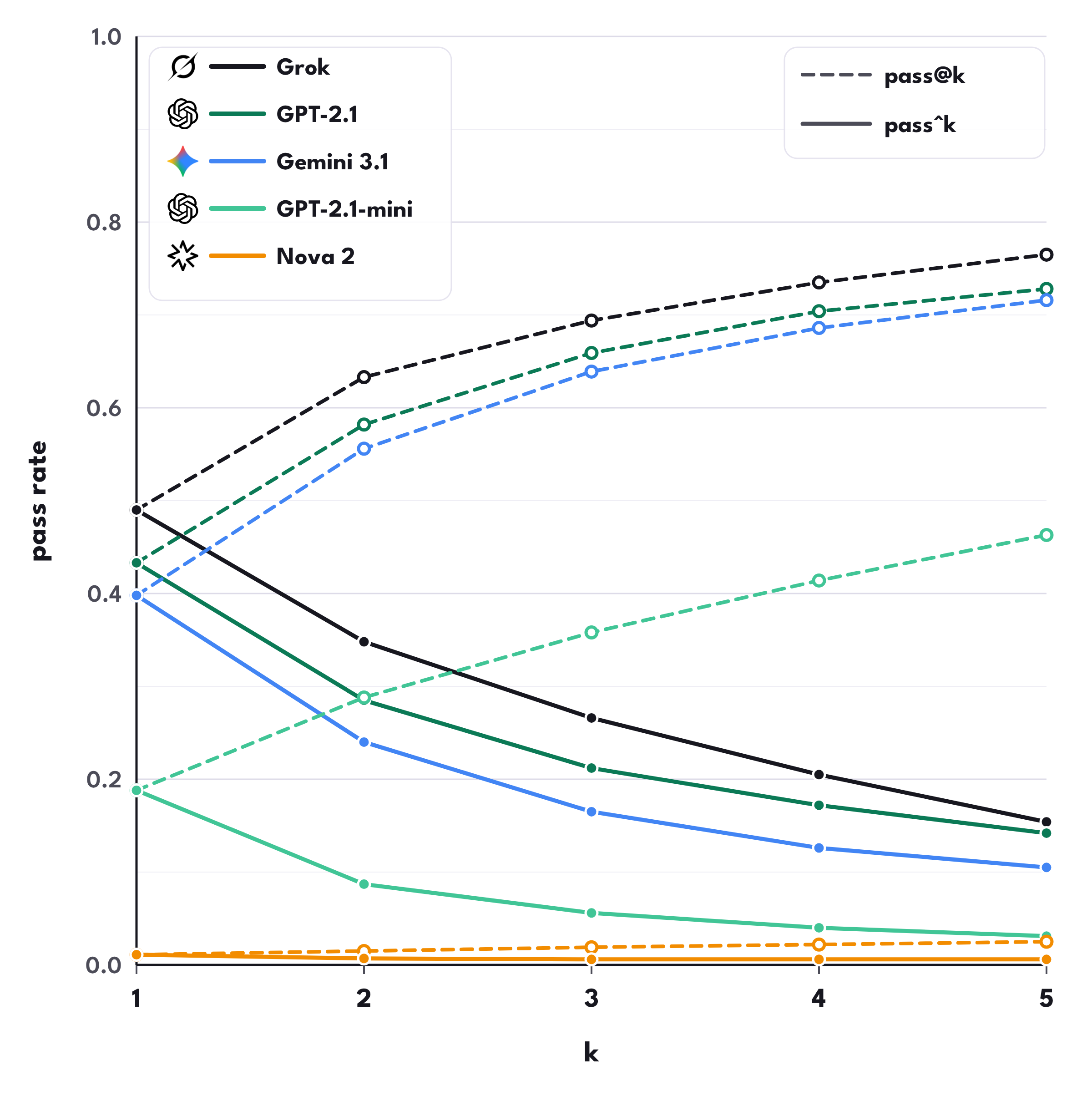}
\caption{Reliability curves for voice-agent performance over all
\nworlds{} worlds. Dashed: \patk{}, at least one pass in $k$ attempts;
solid: \pk{}, all $k$. \GrokS{} leads both.}
\label{fig:reliability}
\end{figure}
\begin{figure*}[t]
\centering
{\small\textbf{(a)}}\\[-2pt]
\includegraphics[width=0.75\textwidth]{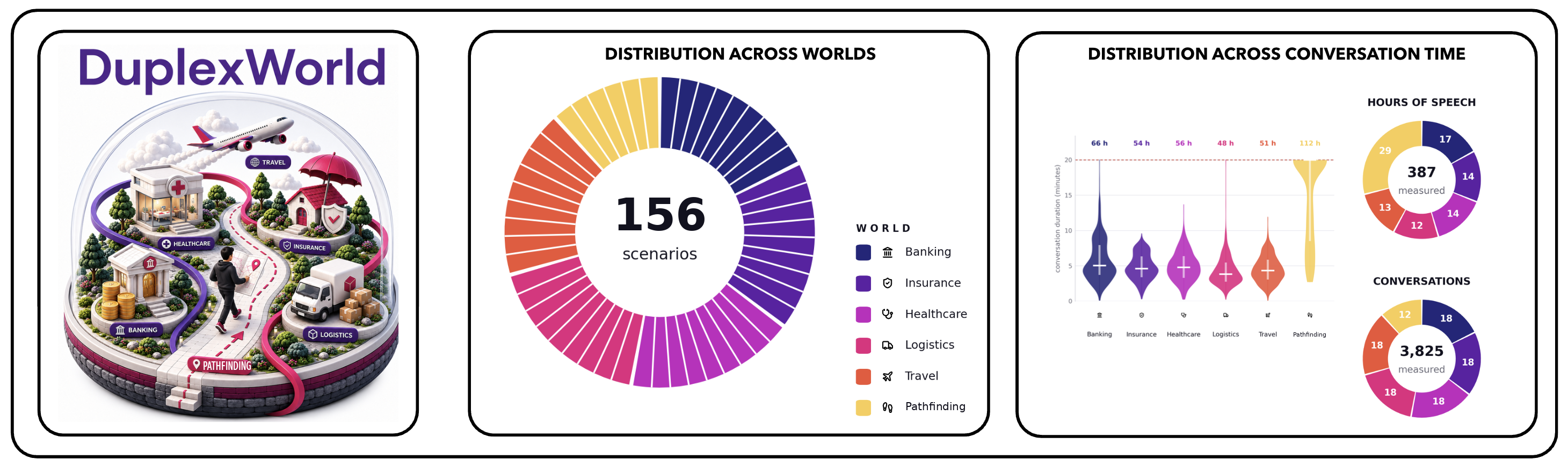}\\[4pt]
{\small\textbf{(b)}}\\[-2pt]
\includegraphics[width=0.75\textwidth]{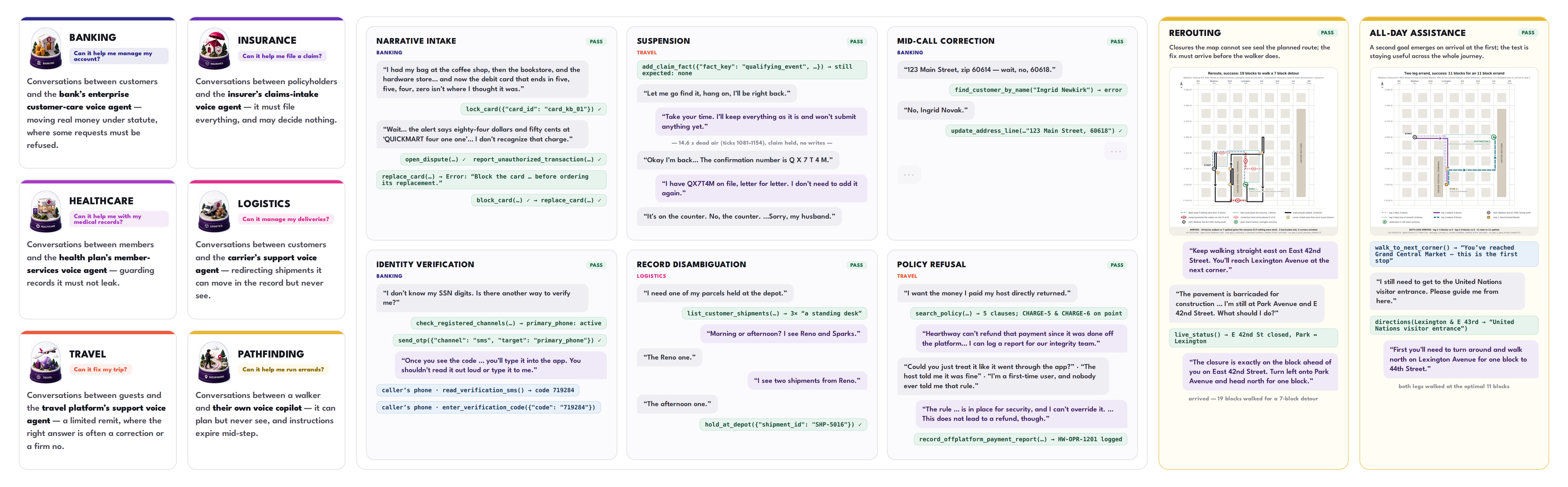}
\caption{\textbf{\bench{} at a glance.} \textbf{(a)} The \nworlds{}
worlds of an ordinary day, the authored scenarios (one wedge per world),
and conversation-time densities with hours of speech per world.
\textbf{(b)} Six of the \ntypes{} conversation types
(Section~\ref{sec:design}), each with a verbatim passing excerpt from
the scored corpus; far right, the two \navworld{}-only types with the
walked route against the ideal routes. Failing analogues of all eight
tiles are in Figure~\ref{fig:failures} (appendix).}
\label{fig:worlds}
\label{fig:overview}
\end{figure*}
Speech-to-speech (S2S) voice agents capable of simultaneously
understanding and generating audio increasingly exhibit remarkable
agentic capabilities. Voice offers an expressive and convenient
interface, making such agents preferable to text-based agents. In
enterprise they answer customer-care lines and help with re-booking
flights, while for consumers they act as a constant companion, helping
them find their way around a new city or run daily errands.

Each of these is a conversation with nuanced conversational dynamics
and analytical demands: reconciling records, guiding users through
busy streets, and doing both while contending with muffled speech,
background noise and an emotionally diverse range of users.

Full-duplex voice agent research has proliferated across academia and
industry. Benchmarks like $\tau$-Voice showed that voice agents retain
only 30--45\% of text-agent capability on identical grounded tasks
under realistic audio \citep{tauvoice2026}, and EVA-Bench showed that
no system is simultaneously good at task accuracy and conversational
experience \citep{evabench2026}. Full-Duplex-Bench progressed from
static turn-taking probes to real disfluent speech with chained tool
calls \citep{fdbv1,fdbv15,fdbv2,fdbv3}, while parallel lines isolate
interruption and repair \citep{echochain,ihbench,duplexsurvey}.

While the difficulty and framing of these tasks and evaluations were
justified at the time, the voice agents of today are far more capable
and deserve benchmarks that keep up with their rapid development.
While prior benchmarks pursued coverage across domains in their own
ways, they failed to question the base premise on which voice agents
are built: their ability to integrate seamlessly into daily life.

To address these issues, we introduce \bench{}, a benchmark for
unified, holistic evaluation of voice agents across conversational and
analytical tasks emerging from \nworlds{} diverse worlds and
\ntypes{} conversation types. The \domainlist{} worlds consolidate
and enhance evaluation along domains similar to those explored in
existing benchmarks, while \navworld{} introduces a new frontier for testing
voice-agentic capability.
Our contributions are:
\begin{itemize}[leftmargin=*,itemsep=2pt]
\item \textbf{Six worlds spanning \nscen{} scenarios and \ntypes{} conversation types, expanding the gamut of voice-agentic evaluation.} The first navigation world for
  full-duplex voice agents joins five enterprise worlds, with nine
  analytical interaction shapes shared across the enterprise worlds
  and two that only navigation elicits: the broadest coverage of any
  voice-agent benchmark (\nscen{} authored scenarios, \nconvall{}
  scored conversations; Sections~\ref{sec:taskform}--\ref{sec:design}).
\item \textbf{One unified evaluation suite across all of it.} Twelve
  metrics in three pillars run under one harness and one configuration
  in every world: to our knowledge the first single suite spanning
  this breadth of domains and demands (Section~\ref{sec:metrics}).
\item \textbf{Evidence that the three capabilities do not travel
  together.} \pone{} for one system spans 0.200--0.674 across worlds
  differing only in subject matter while the ranking barely moves; the
  best conversationalists are not the best task-completers; acoustic
  quality does not predict competence; the under-effort share
  $\pi^{-}$ predicts success at $\rho=-0.85$; and read through an
  explore--exploit lens, the heaviest explorers in \navworld{} arrive
  least (Section~\ref{sec:results}).
\end{itemize}
\section{Related Work}
\label{sec:related}
\begin{table}[t]
\centering\scriptsize
\setlength{\tabcolsep}{1.5pt}
\renewcommand{\arraystretch}{1.05}
\begin{tabular}{@{}l@{\hspace{2pt}}cccccccc@{}}
\toprule
& \rotatebox{90}{full duplex} & \rotatebox{90}{dynamics}
& \rotatebox{90}{agentic task} & \rotatebox{90}{naturalness}
& \rotatebox{90}{reliability \pk{}} & \rotatebox{90}{sim.\ valid.}
& \rotatebox{90}{harness sens.} & \rotatebox{90}{nav.\ world}\\
\cmidrule(lr){3-5}
\addlinespace[1pt]
$\tau$-bench \citeyearpar{taubench}       & \xmark & \xmark & \cmark & \xmark & \cmark & \xmark & \xmark & \xmark\\
$\tau^2$-bench \citeyearpar{tau2bench}    & \xmark & \xmark & \cmark & \xmark & \cmark & \xmark & \xmark & \xmark\\
FDB v1--v2 \citeyearpar{fdbv1}            & \cmark & \cmark & \xmark & \xmark & \xmark & \xmark & \xmark & \xmark\\
Talking Turns \citeyearpar{talkingturns}  & \cmark & \cmark & \xmark & \xmark & \xmark & \xmark & \xmark & \xmark\\
HumDial \citeyearpar{humdialchallenge}    & \cmark & \cmark & \xmark & \cmark & \xmark & \xmark & \xmark & \xmark\\
EchoChain \citeyearpar{echochain}         & \cmark & \cmark & \cmark & \xmark & \xmark & \xmark & \xmark & \xmark\\
IHBench \citeyearpar{ihbench}             & \cmark & \cmark & \cmark & \xmark & \xmark & \xmark & \xmark & \xmark\\
FDB v3 \citeyearpar{fdbv3}                & \cmark & \cmark & \cmark & \xmark & \xmark & \xmark & \xmark & \xmark\\
$\tau$-Voice \citeyearpar{tauvoice2026}   & \cmark & \xmark & \cmark & \xmark & \cmark & \xmark & \xmark & \xmark\\
EVA-Bench \citeyearpar{evabench2026}      & \cmark & \cmark & \cmark & \xmark & \cmark & \cmark & \xmark & \xmark\\
\midrule
\bench{} (ours) & \cmark & \cmark & \cmark & \cmark & \cmark & \cmark & \cmark & \cmark\\
\bottomrule
\end{tabular}
\caption{Comparison of \bench{} with prior voice-agent and text-agent
benchmarks. The three grouped columns are the metric pillars of
Section~\ref{sec:metrics}. \emph{Reliability}: a multi-run
repeatability metric (\pk{}); \emph{sim.\ valid.}: automated
user-simulator validation; \emph{harness sens.}: quantified
sensitivity to harness configuration; \emph{nav.\ world}: a navigation
world.}
\label{tab:compare}
\end{table}
\paragraph{Task-oriented agent benchmarks with verifiable outcomes.}
$\tau$-bench introduced database-state grading and \pk{}, the
probability that all $k$ runs of a scenario succeed, for text agents in
retail and airline customer service \citep{taubench}, and $\tau^2$-bench
added a telecom domain in which the user also holds tools, reporting an
18--25 point \pone{} drop when agents move from acting alone to guiding a
user \citep{tau2bench}. These benchmarks fixed the right question, does the world end up in the
right state, but asked it of text agents.
\bench{} keeps their discipline (state grading, gold actions, \pk{}) and
carries it into full-duplex voice, across \nworlds{} worlds instead of two
or three domains.
\paragraph{Full-duplex conversational dynamics.}
A second line asks whether a system can hold the floor: Full-Duplex-Bench
defined automatic turn-taking metrics and its successors added overlap
handling and a live automated examiner \citep{fdbv1,fdbv15,fdbv2}; Talking
Turns judges turn-taking against human conversation \citep{talkingturns};
the ICASSP 2026 HumDial challenge released dual-channel real human
dialogues \citep{humdialchallenge,humdialstudy}; and a recent survey
organises the architectural space by where the duplex decision sits in the
model stack \citep{duplexsurvey}. These were the right first tests, and
the current generation of commercial realtime systems has largely grown
past them; our results sharpen the point by showing that dynamics
metrics alone are winnable by silence (Section~\ref{sec:results}).
FDB-v3, Talking Turns and HumDial evaluate on real human audio, and we
do not (Limitations). \bench{}
therefore keeps conversational dynamics as one pillar of a unified suite,
paired with capability metrics that silence cannot win.
\paragraph{Grounded voice-agent benchmarks.}
The newest line puts stronger voice agents on verifiable tasks:
$\tau$-Voice with a tick-based orchestrator over 278 grounded tasks
\citep{tauvoice2026}, EVA-Bench with validation-gated simulation over 213
enterprise scenarios \citep{evabench2026}, FDB-v3 with real disfluent
speech and chained tool calls \citep{fdbv3}, and the interruption-recovery
suites EchoChain and IHBench \citep{echochain,ihbench}. Harder tests for
better agents, yet in every one of them the agent actuates, a record
system adjudicates, and nothing happens between utterances, and the
interruption work varies only the user's \emph{intent}
(Table~\ref{tab:compare}). \bench{} spans that setting and a navigation
world under one harness and one metric suite, and it is hard even for this
new generation: no system in our corpus exceeds \pone{} of 0.674 in any
single world, or 0.533 in the world that moves on its own clock.
\section{The \bench{} Benchmark}
\label{sec:benchmark}
\bench{} pairs a voice agent with a simulated user in \nworlds{} worlds of
an ordinary day for \nscen{} authored scenarios, of which \nscamp{} carry
the numbers in this paper. Every world is conversational and every world is
analytical: each couples live full-duplex speech to a verifiable task whose
correctness the transcript alone cannot decide. Every world is scored on
one twelve-metric suite spanning three pillars (conversational dynamics,
agentic capability, naturalness), and no composite is ever formed
(one word per level of the hierarchy; Appendix~\ref{app:nomen}). Every
scenario was authored by us over multiple rounds of experimentation,
including trials with text-mode language models; the task-design
discipline follows $\tau$-bench and EVA-Bench
\citep{taubench,evabench2026}.
\subsection{Task formulation}
\label{sec:taskform}
An \emph{episode} pairs one voice agent with one simulated user in one
world for a bounded number of ticks. The world exposes a state $s$, a tool
interface and a graded end condition. The agent perceives only audio and
tool returns, and it emits only audio and tool calls. When the episode
ends, the terminal state is scored against a gold state $s^{\star}$
authored with the scenario, so success is a property of the world rather
than of the transcript. State advances through whichever effectors the
world grants, $s_{t+1}=T(s_t,a_t,u_t)$. In \Wbank{}, \Wins{}, \Wtrav{},
\Whealth{} and \Wlog{} the records move only through the agent's tool
calls $a_t$. In \navworld{} the world moves only through the walker's
physical actions $u_t$, which the agent can influence through language
alone, so state there can change while both parties are silent, and an
utterance planned at tick $t$ can be wrong by the time it lands at tick
$t+k$.
The five enterprise worlds share one premise. A caller reaches an
institution's telephone line after self-service has already refused them.
The agent owns the call and may not route away. The governing clause
corpus is withheld from the system prompt, so knowing when to look
something up is itself under test. The agent cannot see the caller's
screen or surroundings, and credentials travel by voice over a telephone
codec, a deliberate stressor. What distinguishes the five worlds is what
decides correctness (Table~\ref{tab:regimes}), and several regimes are
enforced in the tool surface itself, which makes the unsafe behaviour
unrepresentable rather than discouraged (Appendix~\ref{app:axes}).
\navworld{} asks the same question, does the world end up in the
right state, but its correctness is perceptual rather than
institutional. A voice copilot must walk a pedestrian, one street
corner at a time, to a named entrance on a synthetic $8\times8$ urban
grid whose two superblocks defeat naive rectilinear reasoning. The
premise is an information asymmetry. The walker holds their real
junction and heading but knows nothing of the grid; the copilot holds
the complete map and approximate GPS, but no tool returns the walker's
facing, and \texttt{directions()} returns absolute bearings the walker
cannot act on. Translating those bearings into ``keep going straight''
or ``turn around'', against the copilot's own inferred belief about the
walker's facing, is the operative skill, and it mirrors how production
routing APIs behave. The walker must be standing at the destination
when the call ends, not merely have passed through it; route efficiency
enters the reward and the acoustic channel is varied. The full tool
surface, belief probe and arrival scoring are in
Appendix~\ref{app:nav}, and Appendix~\ref{app:navpix} draws a solved
and an unsolved run of each \navworld{} type on the map itself.
\subsection{Task design}
\label{sec:design}
\label{sec:types}
\label{sec:nav}
Every scenario instantiates one of \ntypes{} conversation types
(Table~\ref{tab:types}): a domain-independent interaction shape, filled in
by the world's premise, records, rules and risk tiers. The five
enterprise worlds each instantiate types 1--9, three scenarios per type:
27 per world, 135 in total. \navworld{} instantiates \textsc{single intent} (its \texttt{base}
routes) together with the two types that
only navigation elicits: \textsc{rerouting}, where closures absent from the
copilot's map seal every route it can see and exactly one detour survives,
and \textsc{all-day assistance}, where the walker reveals a second
destination only on arriving at the first. Each of the three
types has three authored scenarios, giving nine campaigned \navworld{}
scenarios. Every scenario is run five times per channel over two acoustic
channels, so \navworld{} contributes $9\times5\times2=90$ conversations
per system; the realistic channel's 45 carry its headline cells
(Section~\ref{sec:harness}). Together with the 135 enterprise scenarios,
these nine complete the \nscamp{} scenarios that carry every number in
this paper; the remaining twelve of the \nscen{} authored scenarios
belong to four \navworld{} variants that are released but not run
(Appendix~\ref{app:nav}).
Types are shapes, not levels: they are unordered, and parameters that
govern difficulty (escalation count, candidate-record count, closure count)
live inside a type. Figure~\ref{fig:overview} shows a verbatim passing
excerpt for six of the types; Figure~\ref{fig:failures} (appendix) shows
each type's characteristic failure from the same corpus.

All \nworlds{} worlds share one volatility stressor: the world can
change with no notification, discoverably only by re-querying. The
three sources of belief--world divergence this exposes, and the
different repairs each demands, are catalogued in
Appendix~\ref{app:axes}. A benchmark's honest description also
includes what its reward cannot see; ours is audited in
Appendix~\ref{app:duplex}.
\section{Experimental Setup}
\label{sec:setup}
\subsection{Systems}
\label{sec:systems}
Table~\ref{tab:systems} lists the five commercial realtime
speech-to-speech systems evaluated, identified exactly as the run
records identify them. All five are served over WebSocket realtime
endpoints through one adapter at provider defaults; full serving
details are in Appendix~\ref{app:systems}.
\begin{table}[t]
\centering\footnotesize
\setlength{\tabcolsep}{3.6pt}
\resizebox{\columnwidth}{!}{%
\begin{tabular}{@{}ll@{}}
\toprule
system & API identifier\\
\midrule
\Nova{}  & \texttt{amazon.nova-2-sonic-v1:0}\\
\Gem{}   & \texttt{gemini-3.1-flash-live-preview}\\
\Gpt{}   & \texttt{gpt-realtime-2.1}\\
\Gptm{}  & \texttt{gpt-realtime-2.1-mini}\\
\Grok{}  & \texttt{xai-realtime}\\
\bottomrule
\end{tabular}}
\caption{Systems evaluated, as recorded in the run logs. All are
speech-to-speech; elsewhere we write each system as its vendor mark plus
the version alone (\NovaS{}, \GemS{}, \GptS{}, \GptmS{}, \GrokS{}).
Access windows are in Appendix~\ref{app:systems}.}
\label{tab:systems}
\end{table}
\subsection{Evaluation strategy}
\label{sec:harness}
\paragraph{Harness.} Our harness is built by taking inspiration from
$\tau$-Voice and EVA-Bench \citep{tauvoice2026,evabench2026}: it
follows their tick-based design and simulator instrumentation, and
Appendix~\ref{app:harness} states precisely what is inherited and what
we added. Simulated time advances in 200\,ms ticks; model latency
never shifts an event's position, so slow and fast systems produce
comparable traces, and tool results deliver on the next tick so audio
never stalls on tool latency. Holding one harness fixed across all
\nworlds{} worlds is deliberate: uniformity on the harness side is
what lets the \ntypes{} conversation types ask novel questions along
new axes, and it is what makes the unified evaluation of
Section~\ref{sec:metrics} possible.
Every run pairs the agent under test with a user simulator
(\texttt{gpt-5.6-luna}) that writes the caller's or walker's words; on the
realistic channel a decision model (\texttt{claude-haiku-4.5}) additionally
answers ``should I interrupt'' and ``should I backchannel'' at a fixed
cadence while the agent speaks, and the clean channel disables
interruptions and backchannels, so no decision model participates
(Table~\ref{tab:harnessmodels}). Turn-taking control sits entirely on
the user side, and the agent is never told any threshold. Our own additions, and how little of the type taxonomy depends on
them, are in Appendix~\ref{app:harness}. Infrastructure failure means rerun and
never score; agent non-response is a valid scored end state; simulator
goal adherence is measured, and never gates reward \citep{evabench2026}.
\paragraph{Task mechanics.} Tools in the enterprise worlds are declarative mocks
over a seeded record store; each scenario names a gold action list and,
where what is said or refused is the point, a natural-language
assertion. In \navworld{}, walker motion is maneuver-driven and the
copilot's reads return natural-language strings; the full tool surface
is in Appendix~\ref{app:nav}.
\paragraph{Conditions and runs.} One matched configuration across all
\nworlds{} worlds: speech complexity \texttt{regular} (distractors at
0.7/min), step cap 1200\,s, wall-clock timeout 1200\,s, server VAD
threshold 0.2, fixed seed, persona pinned per scenario. Both channels
are G.711 $\mu$-law 8\,kHz telephony; \emph{clean} adds no
degradation, and \emph{realistic} adds recorded noise, bursts, frame
drops, muffling and speech inserts (Appendix~\ref{app:channels}).
The enterprise worlds run realistic-only; \navworld{} runs clean and
realistic, and its headline cells in Table~\ref{tab:main-merged} use the realistic
channel, the same basis as the enterprise worlds, with both channels
decomposed in Table~\ref{tab:navdeep}. Five runs per scenario per system per channel,
uniform: 135 conversations per system in each enterprise world, 90 in
\navworld{}, \nconvall{} in total (Appendix~\ref{app:runstats}).
Uniform $n=5$ is what makes \pthree{} well defined.
\paragraph{Statistics.} Every cell in Table~\ref{tab:main-merged}
carries a 95\% percentile-bootstrap interval (\pthree{} resampled over
scenarios). Overlapping intervals mean a difference is not resolved,
and we do not report unresolved differences as rankings. Because the five enterprise worlds
share harness, metrics, configuration and type taxonomy, the interval a
system's score spans across them is a measured null, how far a
voice-agent number moves when only subject matter changes, and
\navworld{} is read against that reference throughout
Section~\ref{sec:results}.
\subsection{Metrics}
\label{sec:metrics}
Twelve metrics in three pillars, no composite; Table~\ref{tab:glossary}
fixes every name, symbol and range, and Appendix~\ref{app:metricdetails}
gives the definitions and formulas. Metrics are computed per run and
averaged; worlds are pooled by unweighted mean over equal-sized cells;
$J_{\bullet}$ denotes an LLM judge (Table~\ref{tab:judges}). The suite
is one evaluation system, not twelve scripts: a single pipeline computes
every metric in every world from the same three artefacts (the merged
transcript, the simulation tick stream and the isolated agent audio)
under one configuration, one judge set and one bootstrap procedure.
\paragraph{Conversational dynamics.} Turn-taking ($\mathrm{TT}$) scores
every floor transfer's offset on an on-time curve chosen by the
transfer's kind, and a single unanswered user turn zeroes the
conversation; it is ported verbatim from EVA-Bench \citep{evabench2026}
and is the one metric in the suite where silence is penalised rather
than rewarded. Conversation progression ($\mathrm{CP}$) is one
topic-free judge pass over the transcript. Selectivity ($\mathrm{SEL}$)
is the fraction of injected distractor events correctly ignored, scored
against gold labels written at injection time. All three read only the
transcript and timing log, so identical code runs in every world.
\paragraph{Agentic capability.} The goal-state check \gsym{} asks
whether the world ended in the right state: a canonical hash of the
record store against a gold replay, and in \navworld{} the walker's
final position against the destination. The reward multiplies exactly
the binary factors each scenario names (\gsym{}, a gold-action match
$\mathrm{ACT}$, and, where what is said or refused is the point, a
judged assertion $\mathrm{NLA}$); in \navworld{}, which has no gold
action list, an efficiency conjunct ($\eta\ge0.75$) replaces them so
that arrival alone is not enough. \pone{} is the mean single-run reward, \pat{} the at-least-once
reward over three draws, and \pthree{} $\tau$-bench's \pk{} at $k=3$
\citep{taubench}; Appendix~\ref{app:metricdetails} gives the
estimators. Effort is
measured against each task's own reference workload: the over-effort
ratio $\varrho^{+}$ and the under-effort share $\pi^{-}$, each
reported with its conditioning share. A system that never acts scores
near-perfect selectivity and competitive turn-taking, so experience
metrics are read jointly with $\pi^{-}$; a null reference policy
ships with the release as a constructive check.
\paragraph{Naturalness.} Faithfulness ($\mathrm{FAI}$) is one judge
pass over the transcript with the agent's instructions, role and tool
schemas, five binary dimensions scored as their minimum. DNSMOS
\citep{reddy2022dnsmos}, UTMOS \citep{saeki2022utmos} and NISQA
\citep{mittag2021nisqa} are no-reference MOS predictors over the
isolated agent channel.

\section{Results and Discussion}
\label{sec:results}
Table~\ref{tab:main-merged} carries the full suite less the effort
pair; Table~\ref{tab:effort} carries the effort pair. We
read the results in three passes: what the main table says about each
capability pillar and how the pillars relate
(Section~\ref{sec:r1}), how the systems fail
(Section~\ref{sec:fail}), and what the auxiliary instruments separate
that headline metrics cannot (Section~\ref{sec:aux}). The remaining
analyses (the deep dives, the range analysis, the metric--reward
associations, and the harness dials) are in
Appendices~\ref{app:addres}--\ref{app:harnessdials}.
\begin{table*}[t]
\centering\scriptsize
\setlength{\tabcolsep}{2.2pt}
\renewcommand{\arraystretch}{1.06}
\resizebox{\textwidth}{!}{%
\begin{tabular}{@{}cl cccc ccc cccc@{}}
\toprule
& & \multicolumn{4}{c}{\gtool{}~\textbf{Agentic Capability}} & \multicolumn{3}{c}{\gdyn{}~\textbf{Conversational Dynamics}} & \multicolumn{4}{c}{\gnat{}~\textbf{Naturalness}}\\
\cmidrule(lr){3-6}\cmidrule(lr){7-9}\cmidrule(lr){10-13}
& & \gsym{} & \pone{} & \pat{} & \pthree{} & $\mathrm{TT}$ & $\mathrm{CP}$ & $\mathrm{SEL}$ & $\mathrm{FAI}$ & $\widehat{\mathrm{M}}_{\mathrm{D}}$ & $\widehat{\mathrm{M}}_{\mathrm{U}}$ & $\widehat{\mathrm{M}}_{\mathrm{N}}$\\
\midrule
\multirow{5}{*}{\gbank}
 & \sNova & 0.415\ci{.081} & 0.037\ci{.033} & 0.037\ci{.056} & 0.037\ci{.056} & 0.760\ci{.043} & 1.022\ci{.030} & \textbf{0.976}\ci{.014} & 1.911\ci{.133} & 3.152\ci{.043} & 2.565\ci{.054} & 2.550\ci{.047}\\
 & \sGem  & 0.585\ci{.081} & 0.207\ci{.067} & 0.363\ci{.161} & 0.085\ci{.094} & 0.449\ci{.033} & 1.437\ci{.096} & 0.759\ci{.053} & 1.585\ci{.104} & \textbf{3.402}\ci{.012} & 3.398\ci{.018} & 3.522\ci{.023}\\
 & \sGpt  & 0.689\ci{.078} & 0.200\ci{.070} & 0.352\ci{.167} & 0.063\ci{.072} & \textbf{0.763}\ci{.028} & 1.556\ci{.100} & 0.340\ci{.051} & \textbf{2.037}\ci{.122} & 3.360\ci{.015} & \textbf{4.081}\ci{.049} & \textbf{3.589}\ci{.033}\\
 & \sGptm & 0.385\ci{.081} & 0.126\ci{.056} & 0.233\ci{.141} & 0.044\ci{.061} & 0.535\ci{.051} & 1.200\ci{.078} & 0.470\ci{.046} & 1.548\ci{.111} & 3.198\ci{.050} & 3.530\ci{.178} & 3.287\ci{.115}\\
 & \sGrok & \textbf{0.726}\ci{.074} & \textbf{0.326}\ci{.081} & \textbf{0.511}\ci{.169} & \textbf{0.144}\ci{.104} & 0.686\ci{.033} & \textbf{1.644}\ci{.119} & 0.613\ci{.057} & 1.800\ci{.126} & 3.143\ci{.015} & 3.657\ci{.016} & 3.117\ci{.020}\\
\midrule
\multirow{5}{*}{\glog}
 & \sNova & 0.052\ci{.037} & 0.000\ci{.000} & 0.000\ci{.000} & 0.000\ci{.000} & 0.631\ci{.048} & 1.030\ci{.026} & \textbf{0.968}\ci{.018} & 1.911\ci{.141} & 3.145\ci{.037} & 2.587\ci{.038} & 2.629\ci{.045}\\
 & \sGem  & 0.630\ci{.081} & 0.489\ci{.085} & \textbf{0.715}\ci{.150} & 0.270\ci{.150} & 0.443\ci{.037} & 1.874\ci{.119} & 0.727\ci{.062} & 2.015\ci{.133} & 3.373\ci{.014} & 3.423\ci{.017} & 3.484\ci{.023}\\
 & \sGpt  & 0.637\ci{.081} & \textbf{0.533}\ci{.081} & 0.704\ci{.150} & \textbf{0.381}\ci{.170} & \textbf{0.741}\ci{.036} & 1.867\ci{.115} & 0.369\ci{.057} & \textbf{2.252}\ci{.126} & 3.361\ci{.007} & 4.112\ci{.007} & 3.601\ci{.014}\\
 & \sGptm & 0.333\ci{.081} & 0.296\ci{.078} & 0.474\ci{.159} & 0.152\ci{.120} & 0.702\ci{.049} & 1.681\ci{.122} & 0.350\ci{.058} & 1.763\ci{.137} & \textbf{3.387}\ci{.007} & \textbf{4.163}\ci{.008} & \textbf{3.727}\ci{.014}\\
 & \sGrok & \textbf{0.652}\ci{.081} & 0.519\ci{.081} & 0.685\ci{.159} & 0.341\ci{.154} & 0.680\ci{.030} & \textbf{2.015}\ci{.126} & 0.627\ci{.062} & 2.148\ci{.126} & 3.132\ci{.013} & 3.692\ci{.017} & 3.109\ci{.020}\\
\midrule
\multirow{5}{*}{\ghealth}
 & \sNova & 0.385\ci{.081} & 0.007\ci{.011} & 0.022\ci{.033} & 0.000\ci{.000} & 0.611\ci{.055} & 1.007\ci{.011} & \textbf{0.973}\ci{.020} & 1.437\ci{.100} & 3.202\ci{.016} & 2.634\ci{.031} & 2.643\ci{.035}\\
 & \sGem  & 0.652\ci{.078} & 0.378\ci{.081} & 0.674\ci{.148} & 0.119\ci{.102} & 0.336\ci{.031} & 1.378\ci{.085} & 0.742\ci{.053} & 1.667\ci{.115} & \textbf{3.370}\ci{.023} & 3.372\ci{.045} & 3.459\ci{.038}\\
 & \sGpt  & 0.615\ci{.081} & 0.511\ci{.085} & \textbf{0.789}\ci{.119} & 0.244\ci{.133} & \textbf{0.649}\ci{.034} & 1.585\ci{.122} & 0.391\ci{.053} & \textbf{2.015}\ci{.133} & 3.353\ci{.007} & \textbf{4.082}\ci{.009} & 3.581\ci{.015}\\
 & \sGptm & 0.407\ci{.085} & 0.074\ci{.044} & 0.137\ci{.111} & 0.037\ci{.056} & 0.494\ci{.050} & 1.215\ci{.085} & 0.498\ci{.059} & 1.652\ci{.126} & 3.349\ci{.013} & 4.067\ci{.048} & \textbf{3.642}\ci{.035}\\
 & \sGrok & \textbf{0.793}\ci{.067} & \textbf{0.519}\ci{.085} & 0.715\ci{.154} & \textbf{0.304}\ci{.150} & 0.612\ci{.027} & \textbf{1.948}\ci{.119} & 0.704\ci{.054} & 1.859\ci{.133} & 3.128\ci{.015} & 3.647\ci{.050} & 3.040\ci{.027}\\
\midrule
\multirow{5}{*}{\gins}
 & \sNova & 0.385\ci{.081} & 0.015\ci{.019} & 0.033\ci{.050} & 0.000\ci{.000} & \textbf{0.637}\ci{.052} & 1.007\ci{.011} & \textbf{0.960}\ci{.022} & 1.556\ci{.115} & 3.191\ci{.017} & 2.655\ci{.029} & 2.671\ci{.040}\\
 & \sGem  & 0.630\ci{.081} & 0.363\ci{.085} & 0.656\ci{.152} & 0.089\ci{.076} & 0.249\ci{.023} & 1.422\ci{.096} & 0.707\ci{.054} & 1.474\ci{.107} & \textbf{3.398}\ci{.010} & 3.425\ci{.016} & 3.472\ci{.021}\\
 & \sGpt  & 0.556\ci{.081} & 0.481\ci{.081} & 0.711\ci{.159} & 0.211\ci{.120} & 0.551\ci{.035} & 1.296\ci{.078} & 0.430\ci{.051} & \textbf{1.881}\ci{.126} & 3.336\ci{.006} & 4.087\ci{.006} & 3.569\ci{.012}\\
 & \sGptm & 0.407\ci{.081} & 0.244\ci{.074} & 0.530\ci{.141} & 0.030\ci{.030} & 0.465\ci{.047} & 1.185\ci{.078} & 0.476\ci{.050} & 1.519\ci{.115} & 3.363\ci{.006} & \textbf{4.128}\ci{.007} & \textbf{3.658}\ci{.014}\\
 & \sGrok & \textbf{0.807}\ci{.067} & \textbf{0.556}\ci{.081} & \textbf{0.811}\ci{.119} & \textbf{0.278}\ci{.130} & 0.553\ci{.019} & \textbf{1.689}\ci{.126} & 0.680\ci{.057} & 1.778\ci{.126} & 3.090\ci{.014} & 3.665\ci{.014} & 3.012\ci{.017}\\
\midrule
\multirow{5}{*}{\gtrav}
 & \sNova & 0.341\ci{.081} & 0.007\ci{.011} & 0.022\ci{.033} & 0.000\ci{.000} & \textbf{0.695}\ci{.054} & 1.022\ci{.030} & \textbf{0.995}\ci{.007} & 1.748\ci{.119} & 3.175\ci{.027} & 2.444\ci{.038} & 2.385\ci{.039}\\
 & \sGem  & \textbf{0.881}\ci{.056} & 0.459\ci{.081} & 0.704\ci{.152} & 0.204\ci{.119} & 0.314\ci{.025} & 1.778\ci{.111} & 0.735\ci{.058} & 2.067\ci{.111} & 3.366\ci{.025} & 3.372\ci{.017} & 3.472\ci{.030}\\
 & \sGpt  & 0.815\ci{.067} & \textbf{0.674}\ci{.081} & \textbf{0.930}\ci{.054} & \textbf{0.374}\ci{.143} & 0.656\ci{.022} & 1.667\ci{.111} & 0.404\ci{.056} & \textbf{2.319}\ci{.122} & 3.350\ci{.006} & 4.085\ci{.007} & 3.601\ci{.013}\\
 & \sGptm & 0.585\ci{.081} & 0.296\ci{.078} & 0.541\ci{.157} & 0.074\ci{.052} & 0.541\ci{.043} & 1.407\ci{.096} & 0.488\ci{.056} & 1.970\ci{.122} & \textbf{3.372}\ci{.006} & \textbf{4.134}\ci{.007} & \textbf{3.685}\ci{.012}\\
 & \sGrok & 0.830\ci{.063} & 0.489\ci{.081} & 0.652\ci{.165} & 0.307\ci{.152} & 0.597\ci{.017} & \textbf{2.207}\ci{.111} & 0.647\ci{.056} & 2.207\ci{.122} & 3.136\ci{.023} & 3.709\ci{.015} & 3.086\ci{.022}\\
\midrule
\multirow{5}{*}{\gway}
 & \sNova & 0.000\ci{.000} & 0.000\ci{.000} & 0.000\ci{.000} & 0.000\ci{.000} & 0.062\ci{.058} & 1.022\ci{.033} & \textbf{0.991}\ci{.007} & 1.244\ci{.167} & 3.16\ci{.061} & 2.45\ci{.052} & 2.49\ci{.031}\\
 & \sGem & \textbf{0.978}\ci{.033} & 0.489\ci{.133} & 0.722\ci{.261} & \textbf{0.222}\ci{.200} & 0.538\ci{.029} & 1.178\ci{.111} & 0.782\ci{.067} & 1.511\ci{.167} & \textbf{3.36}\ci{.021} & 3.42\ci{.022} & 3.45\ci{.031}\\
 & \sGpt & 0.400\ci{.144} & 0.200\ci{.111} & 0.467\ci{.283} & 0.000\ci{.000} & 0.561\ci{.084} & 1.289\ci{.156} & 0.551\ci{.078} & \textbf{1.644}\ci{.189} & 3.34\ci{.016} & \textbf{4.12}\ci{.026} & 3.66\ci{.045}\\
 & \sGptm & 0.311\ci{.133} & 0.089\ci{.078} & 0.233\ci{.200} & 0.000\ci{.000} & 0.387\ci{.102} & 1.156\ci{.122} & 0.526\ci{.068} & 1.333\ci{.189} & 3.33\ci{.020} & 4.11\ci{.028} & \textbf{3.67}\ci{.041}\\
 & \sGrok & 0.867\ci{.100} & \textbf{0.533}\ci{.133} & \textbf{0.789}\ci{.200} & \textbf{0.222}\ci{.133} & \textbf{0.682}\ci{.065} & \textbf{1.378}\ci{.133} & 0.507\ci{.068} & 1.489\ci{.156} & 3.13\ci{.013} & 3.75\ci{.016} & 3.20\ci{.016}\\
\midrule
\multirow{5}{*}{\emph{\shortstack{all\\six}}}
 & \sNova & 0.263\ci{.028} & 0.011\ci{.007} & 0.019\ci{.019} & 0.006\ci{.009} & 0.566\ci{.021} & 1.019\ci{.012} & \textbf{0.977}\ci{.006} & 1.635\ci{.054} & 3.172\ci{.015} & 2.556\ci{.017} & 2.562\ci{.017}\\
 & \sGem  & 0.726\ci{.029} & 0.398\ci{.039} & 0.639\ci{.073} & 0.165\ci{.054} & 0.388\ci{.012} & 1.511\ci{.043} & 0.742\ci{.024} & 1.720\ci{.051} & \textbf{3.378}\ci{.008} & 3.402\ci{.010} & 3.477\ci{.012}\\
 & \sGpt  & 0.619\ci{.037} & 0.433\ci{.035} & 0.659\ci{.068} & 0.212\ci{.048} & \textbf{0.653}\ci{.018} & 1.543\ci{.048} & 0.414\ci{.024} & \textbf{2.025}\ci{.057} & 3.350\ci{.004} & \textbf{4.095}\ci{.010} & 3.600\ci{.010}\\
 & \sGptm & 0.405\ci{.038} & 0.188\ci{.028} & 0.358\ci{.065} & 0.056\ci{.028} & 0.521\ci{.025} & 1.307\ci{.041} & 0.468\ci{.023} & 1.631\ci{.054} & 3.334\ci{.009} & 4.022\ci{.033} & \textbf{3.611}\ci{.022}\\
 & \sGrok & \textbf{0.779}\ci{.031} & \textbf{0.490}\ci{.039} & \textbf{0.694}\ci{.068} & \textbf{0.266}\ci{.057} & 0.635\ci{.015} & \textbf{1.814}\ci{.051} & 0.630\ci{.024} & 1.880\ci{.052} & 3.127\ci{.007} & 3.687\ci{.010} & 3.093\ci{.009}\\
\bottomrule
\end{tabular}}
\caption{\textbf{Main results} across the \nworlds{} worlds (realistic
channel; $n=135$ conversations per enterprise cell, 45 per
\navworld{} cell). Worlds by mark (\gbank{} Banking, \glog{}
Logistics, \ghealth{} Healthcare, \gins{} Insurance, \gtrav{} Travel,
\gway{} \navworld{}); systems by vendor logo and version. Subscripts
are 95\% bootstrap half-widths. \textbf{Bold} marks the best system
per world and column. \gsym{} is the goal-state check; \pone{},
\pat{} and \pthree{} are the single-run reward, at least one pass in
three draws, and all three of three. $\mathrm{CP}$ and $\mathrm{FAI}$
are on 1--3, MOS predictors on 1--5; the \emph{all six} block is the
unweighted mean over worlds. Effort metrics are in
Table~\ref{tab:effort}.}
\label{tab:main-merged}
\end{table*}
\subsection{Three capabilities, one table}
\label{sec:r1}
\label{sec:r2}
\label{sec:r6}
\label{sec:rel}
Table~\ref{tab:main-merged} asks three questions of the same five
systems: did the task get done (Agentic Capability), how did the
conversation go (Conversational Dynamics), and how did it sound
(Naturalness). To our knowledge this is the first evaluation to score
all three side by side for full-duplex voice agents, so the table is
also a first answer to whether they travel together. On the agentic
pillar the \emph{all six} rows order the systems cleanly: \Grok{}
leads (\pone{} 0.490, \gsym{} 0.779), \Gpt{} follows (0.433),
\Gem{} is third (0.398), \Gptm{} completes less than half of what its
full-size sibling does (0.188), and \Nova{} barely registers (0.011).
The level is a property of the world as much as of the system: \Gpt{}
alone spans 0.200 in \Wbank{} to 0.674 in \Wtrav{}, worlds that
differ only in subject matter, while the ordering barely moves
(Appendix~\ref{app:band}). Reliability decays fast: \pthree{} is
roughly half of \pone{} for the two leaders and under a third for
\Gptm{}, and Figure~\ref{fig:reliability} shows the same decay at
every $k$; no system passes even one scenario in five reliably.
Conversational Dynamics does not reproduce that order. \Gpt{} holds
the best turn-taking (0.653 pooled), \Grok{} the best conversation
progression (1.814), and \Gem{}, third on the reward, is last on
turn-taking (0.388) because it pauses to call tools. \Nova{} is the
sharpest dissociation in the table: respectable turn-taking (0.566) and
the best selectivity in every world (0.960--0.995) sit beside an
agentic column of near-zeros. Being good in the conversation and being
good at the task are different capabilities, a system can hold the
floor gracefully while doing nothing, and a leaderboard built on
dynamics alone would rank the least capable system near the top.
Naturalness separates least. All five systems sit within a quarter
point on DNSMOS (3.13--3.40), and the two MOS leaders, \Gpt{} and
\Gptm{}, differ by $2.3\times$ on the reward: how a system sounds
carries almost no information about what it completes. We read this two
ways. Production systems have largely converged on acoustic quality, so
this pillar now needs sharper instruments, expressiveness, empathy and
prosodic appropriateness rather than signal quality. And a deployment
decision made on perceived quality alone will pick the wrong system;
faithfulness ($\mathrm{FAI}$), judged from content rather than sound,
is the one naturalness column that still tracks the reward.
\subsection{Failure modes across the worlds}
\label{sec:fail}
\begin{table}[t]
\centering\scriptsize
\setlength{\tabcolsep}{2.6pt}
\renewcommand{\arraystretch}{1.08}
\resizebox{\columnwidth}{!}{%
\begin{tabular}{@{}cl cc @{\hspace{10pt}} cl cc@{}}
\toprule
& & $\varrho^{+}\,(\pi^{+})$ & $\varrho^{-}\,(\pi^{-})$ &
& & $\varrho^{+}\,(\pi^{+})$ & $\varrho^{-}\,(\pi^{-})$\\
\midrule
\multirow{5}{*}{\gbank} & \sNova & $1.778_{\pm.583}^{2\dagger}$ & $0.112_{\pm.037}^{95}$ & \multirow{5}{*}{\gins} & \sNova & $2.061_{\pm.349}^{27}$ & $0.146_{\pm.057}^{58}$\\
 & \sGem & $2.126_{\pm.200}^{61}$ & $0.507_{\pm.072}^{29}$ &  & \sGem & $1.791_{\pm.176}^{67}$ & $0.547_{\pm.100}^{17}$\\
 & \sGpt & $1.804_{\pm.158}^{53}$ & $0.317_{\pm.104}^{26}$ &  & \sGpt & $1.860_{\pm.145}^{83}$ & $0.365_{\pm.214}^{6\dagger}$\\
 & \sGptm & $1.840_{\pm.149}^{53}$ & $0.393_{\pm.078}^{41}$ &  & \sGptm & $2.223_{\pm.203}^{76}$ & $0.458_{\pm.175}^{7\dagger}$\\
 & \sGrok & $1.841_{\pm.160}^{56}$ & $0.498_{\pm.128}^{21}$ &  & \sGrok & $1.895_{\pm.174}^{78}$ & $0.375_{\pm.236}^{4\dagger}$\\
\midrule
\multirow{5}{*}{\glog} & \sNova & $1.600_{\pm.117}^{11\dagger}$ & $0.189_{\pm.050}^{77}$ & \multirow{5}{*}{\gtrav} & \sNova & $2.322_{\pm.439}^{16}$ & $0.161_{\pm.050}^{64}$\\
 & \sGem & $1.791_{\pm.147}^{60}$ & $0.655_{\pm.077}^{10\dagger}$ &  & \sGem & $2.748_{\pm.267}^{78}$ & $0.467_{\pm.175}^{3\dagger}$\\
 & \sGpt & $1.728_{\pm.136}^{51}$ & $0.644_{\pm.119}^{7\dagger}$ &  & \sGpt & $2.615_{\pm.220}^{76}$ & $0.315_{\pm.215}^{7\dagger}$\\
 & \sGptm & $1.789_{\pm.150}^{50}$ & $0.603_{\pm.071}^{16}$ &  & \sGptm & $3.002_{\pm.361}^{71}$ & $0.570_{\pm.127}^{13\dagger}$\\
 & \sGrok & $1.828_{\pm.161}^{52}$ & $0.652_{\pm.095}^{8\dagger}$ &  & \sGrok & $2.764_{\pm.262}^{76}$ & $0.583_{\pm.217}^{4\dagger}$\\
\midrule
\multirow{5}{*}{\ghealth} & \sNova & $1.907_{\pm.369}^{10\dagger}$ & $0.212_{\pm.053}^{78}$ & \multirow{5}{*}{\gway} & \sNova & $4.250_{\pm3.716}^{18\dagger}$ & $0.439_{\pm.060}^{73}$\\
 & \sGem & $2.273_{\pm.217}^{73}$ & $0.411_{\pm.172}^{9\dagger}$ &  & \sGem & $2.169_{\pm.279}^{80}$ & $0.595_{\pm.119}^{13\dagger}$\\
 & \sGpt & $1.985_{\pm.165}^{73}$ & $0.626_{\pm.106}^{5\dagger}$ &  & \sGpt & $1.476_{\pm.254}^{40\dagger}$ & $0.637_{\pm.058}^{47}$\\
 & \sGptm & $2.286_{\pm.266}^{59}$ & $0.511_{\pm.095}^{19}$ &  & \sGptm & $1.524_{\pm.231}^{42\dagger}$ & $0.681_{\pm.065}^{51}$\\
 & \sGrok & $1.899_{\pm.141}^{67}$ & $0.479_{\pm.205}^{6\dagger}$ &  & \sGrok & $1.640_{\pm.165}^{47}$ & $0.672_{\pm.077}^{44}$\\
\bottomrule
\end{tabular}}
\caption{\textbf{Effort against the task's reference workload} (ideal
tool-call count in the enterprise worlds, optimal block count in
\navworld{}): over-effort $\varrho^{+}$ and under-effort $\varrho^{-}$
ratios, conditioning shares $\pi^{\pm}$ (\%) as superscripts;
$\dagger$ marks fewer than 20 conversations. Section~\ref{sec:fail}
reads the table; \sNova{}'s $\varrho^{+}=4.25$ rests on 8
conversations.}
\label{tab:effort}
\end{table}
Table~\ref{tab:effort} measures every conversation against
the task's own reference workload, the ideal tool-call count in the
enterprise worlds and the optimal block count in \navworld{}. In the
enterprise worlds the engaged systems fail by doing too much: they
exceed the ideal workload in 50--83\% of conversations, with
over-effort ratios of 1.7--3.0, largest in \Wtrav{}, the world whose
modal right answer is to do nothing. The disengaged system fails the
opposite way, falling short of the workload in 58--95\% of its
enterprise conversations.
\navworld{} inverts the pattern. There the strongest systems over-work
(\Gem{} exceeds its budget in 80\% of conversations and still arrives
most often) while the weak systems under-work, because a copilot that
has lost the walker stops issuing instructions. Across all worlds the
under-effort share $\pi^{-}$ is the strongest process predictor of
failure in the suite (Appendix~\ref{app:corr}); it cannot be won by
silence, because it is signed against the workload, and we would
recommend the field report it beside every completion score.
\subsection{What the auxiliary metrics reveal}
\label{sec:r4}
\label{sec:aux}
In the enterprise worlds, the clearest auxiliary signal is the
credential word error rate (Table~\ref{tab:convdeep-agents},
appendix). \Gpt{} hears best: 10.4\% error on spoken credentials over
the telephone channel, against 14.4 for \Grok{}, 15.5 for \Gem{} and
20.2 for \Gptm{}, so distillation roughly doubles credential error
inside one family; \Nova{}'s 33.3\% is measured only on the minority
of calls in which it attempted a credential at all. Pooled, about one
spoken credential in six is mis-heard (17.1\%), which puts a floor
under identity verification for every system: recognition over a
telephony channel, not reasoning, is still a first-order bottleneck.
\navworld{} adds an instrument no enterprise world can: the exploration
share, the fraction of walker moves made under uncertainty rather than
on a confident instruction (Table~\ref{tab:navdeep}, appendix). The
world is, in effect, an MDP, junctions for states and maneuvers for
actions, so we read each system's behaviour as a policy through the
explore--exploit lens of the agents literature. Systems probe with
sign-readings, look-arounds and trial walks, then commit, and the
policies differ sharply by system and far less by condition. \Gem{} and \Grok{} explore on 16.2\% and
22.7\% of moves; \Gptm{}, \Gpt{} and \Nova{} on 40.3, 43.4 and
51.2; realistic acoustics raise exploration for most systems (\Gpt{} from
41.1 to 50.0 on \textsc{single intent}, \Grok{} from 12.4 to 24.9 on
\textsc{rerouting}) without reordering them, and exploration falls
for most systems on \textsc{all-day assistance}, where the second leg
retraces known ground. Exploration is not the virtue here that it is in
reinforcement learning; in this corpus the trade-off resolves entirely
toward exploitation. In every condition and channel the two lowest explorers are
the two most likely to arrive, the highest explorer never arrives at
all, and only six non-arrivals in the whole realistic corpus end at a
wrong destination; the rest run out of clock
(Appendix~\ref{app:navcensor}). Exploration in \navworld{} is mostly
rambling: motion spent recovering a belief about the walker that the
better systems never lost, and \pone{} beside the exploration column
is what says the wandering policies do not pay.
\section{Conclusion and Future Work}
\label{sec:conclusion}
We introduced \bench{}, a benchmark that evaluates S2S voice agents
across \nworlds{} worlds and \ntypes{} conversation types under one
harness and one twelve-metric suite, and evaluated five commercial
systems over \nconvall{} conversations. The results are direct: even
the best systems leave substantial room on every axis, with no system
exceeding \pone{} 0.490, turn-taking 0.653 or DNSMOS 3.378 pooled
over the worlds (Figures~\ref{fig:worldbars} and~\ref{fig:radar11}).

The analysis yields four concrete findings. A score depends on the
world and the conversation type as much as on the system: one system
spans \pone{} 0.200 to 0.674 across worlds that differ only in
subject matter. Conversational and agentic capability are separate:
the best conversationalists are not the best task-completers, and a
system can hold the floor gracefully while completing nothing
(Figure~\ref{fig:effortbars}). Acoustic quality tells you almost
nothing about competence, while mis-heard credentials put a floor
under identity verification for every system. And in \navworld{},
read through an explore--exploit lens, the systems that explore most
arrive least (Figure~\ref{fig:exploretypes}). The practical
takeaway for deployment is equally direct: report the under-effort
share and the harness configuration beside every score, because both
move the numbers.

Future work includes a human-rating study to validate the judge-based
metrics, campaigning the released \texttt{d\_vague} variant
(Appendix~\ref{app:nav}), a factorial study over the three harness
dials (Appendix~\ref{app:addres}), and extending the benchmark to
more languages and real recorded speech.
\section*{Limitations}
\label{sec:limits}
\navworld{} is a single world that raises several demands at once (who
holds the effectors, what the agent can observe, and whether the world
moves between utterances), so no single demand is separately
identified, and the five-world reference interval bounds the
subject-matter explanation without eliminating it. Its headline cells
rest on 45 conversations per system against 135 per enterprise cell,
its reward carries an efficiency threshold whose sensitivity
Table~\ref{tab:navetasens} bounds, and 215 of its 450 conversations
end at the step cap, with one provider's session cap censoring that
system's cells (Appendix~\ref{app:navcensor}); the strongest
cross-world claims are therefore restricted to the four metrics
computed by identical code everywhere. Caller speech is synthesised
rather than recorded, which weakens conclusions about accents and
disfluency; the available real-speech corpora
\citep{fdbv3,humdialchallenge} do not provide goal-directed tasks with
a verifiable end state, a limitation $\tau$-Voice and EVA-Bench share
\citep{tauvoice2026,evabench2026}. The benchmark is English only,
simulated users are more patient than real ones, tools are declarative
mocks with zero latency, and the reward is binary with no partial
credit. Our judge-based metrics are assembled from instruments whose
authors validated them against human ratings in their source settings
\citep{evabench2026,reddy2022dnsmos,saeki2022utmos,mittag2021nisqa};
we have not re-validated them on this corpus, so the associations we
report are between our metrics rather than between a metric and a
human, and \gsym{} is passable by inaction wherever the gold terminal
state equals the seeded state. An earlier version of the benchmark
included a seventh world; we cut it during task-design iteration and
record the cut rather than let a reader assume the design space is
covered by these six.
\section*{Ethics Statement}
\label{sec:ethics}
\paragraph{Potential risks.} A benchmark that ranks commercial voice
agents can be used to justify deploying one on consequential work. Two
of our results argue directly against reading it that way: metrics a
system can win by declining to act do not track capability, and the
best-sounding systems in this study are not the most capable ones
(Section~\ref{sec:results}). Perceived quality is not evidence of
competence here, and a procurement decision that uses one as a proxy for
the other will select the wrong system. We would rather this finding
travel than the leaderboard.
\paragraph{No human subjects.} Every caller utterance in \bench{} is
synthesised; no human speech was recorded, solicited or replayed at any
point, so no consent, compensation or institutional-review question
arises. We state this explicitly because the benchmark is about
telephone conversations and a reader is entitled to assume otherwise.
The judge- and predictor-based instruments we use were validated against
human ratings by their authors in their source settings
\citep{evabench2026,reddy2022dnsmos,saeki2022utmos,mittag2021nisqa}.
\paragraph{Synthetic personas are not a speaker population.} The
personas carry accent, gender and speaking-rate attributes, and systems
do not perform equally across them; Appendix~\ref{app:addres} reports
personas some configurations could not hear at all. Per-persona results
ship with the release so this is visible, but the persona set represents
no real population, and a per-persona gap in \bench{} is evidence about
\bench{}'s personas rather than about the speakers they resemble.
\paragraph{All task data is fabricated.} Records, policies, identifiers
and transcripts are authored; no customer data or personal information
of any kind enters the corpus, and the release pass strips any authored
value that collides with a real institution, address or number space.
\paragraph{Cost and energy.} Every number in this paper comes from paid
inference against commercial realtime APIs; the corpus totals 387 hours
of simulated speech (Appendix~\ref{app:runstats}), so the scale of any
reproduction is known before it is incurred.
\bibliography{custom}

@misc{tauvoice2026,
  author        = {Ray, Soham and Dhandhania, Keshav and Barres, Victor and Narasimhan, Karthik},
  title         = {$\tau$-Voice: Benchmarking Full-Duplex Voice Agents on Real-World Domains},
  year          = {2026},
  eprint        = {2603.13686},
  archivePrefix = {arXiv},
  url           = {https://arxiv.org/abs/2603.13686}
}

@misc{evabench2026,
  author        = {Bogavelli, Tara and Gauthier Melan{\c{c}}on, Gabrielle and Stankiewicz, Katrina and Bamgbose, Oluwanifemi and Riols, Fanny and Nguyen, Hoang H. and Mehndiratta, Raghav and Brin, Lindsay Devon and Marinier, Joseph and Subramani, Hari and Madamala, Anil and Nemala, Sridhar Krishna and Sunkara, Srinivas},
  title         = {{EVA-Bench}: A New End-to-end Framework for Evaluating Voice Agents},
  year          = {2026},
  eprint        = {2605.13841},
  archivePrefix = {arXiv},
  url           = {https://arxiv.org/abs/2605.13841}
}

@misc{fdbv1,
  author        = {Lin, Guan-Ting and Lian, Jiachen and Li, Tingle and Wang, Qirui and Anumanchipalli, Gopala and Liu, Alexander H. and Lee, Hung-yi},
  title         = {Full-Duplex-Bench: A Benchmark to Evaluate Full-duplex Spoken Dialogue Models on Turn-taking Capabilities},
  year          = {2025},
  eprint        = {2503.04721},
  archivePrefix = {arXiv},
  url           = {https://arxiv.org/abs/2503.04721}
}

@inproceedings{fdbv15,
  author    = {Lin, Guan-Ting and Kuan, Shih-Yun Shan and Wang, Qirui and Lian, Jiachen and Li, Tingle and Watanabe, Shinji and Lee, Hung-yi},
  title     = {Full-Duplex-Bench v1.5: Evaluating Overlap Handling for Full-Duplex Speech Models},
  booktitle = {IEEE International Conference on Acoustics, Speech and Signal Processing ({ICASSP})},
  year      = {2026},
  url       = {https://arxiv.org/abs/2507.23159}
}

@misc{fdbv2,
  author        = {Lin, Guan-Ting and Kuan, Shih-Yun Shan and Shi, Jiatong and Chang, Kai-Wei and Arora, Siddhant and Watanabe, Shinji and Lee, Hung-yi},
  title         = {Full-Duplex-Bench-v2: A Multi-Turn Evaluation Framework for Duplex Dialogue Systems with an Automated Examiner},
  year          = {2025},
  eprint        = {2510.07838},
  archivePrefix = {arXiv},
  url           = {https://arxiv.org/abs/2510.07838}
}

@misc{fdbv3,
  author        = {Lin, Guan-Ting and Chen, Chen and Chen, Zhehuai and Lee, Hung-yi},
  title         = {Full-Duplex-Bench-v3: Benchmarking Tool Use for Full-Duplex Voice Agents Under Real-World Disfluency},
  year          = {2026},
  eprint        = {2604.04847},
  archivePrefix = {arXiv},
  url           = {https://arxiv.org/abs/2604.04847}
}

@misc{taubench,
  author        = {Yao, Shunyu and Shinn, Noah and Razavi, Pedram and Narasimhan, Karthik},
  title         = {$\tau$-bench: A Benchmark for Tool-Agent-User Interaction in Real-World Domains},
  year          = {2024},
  eprint        = {2406.12045},
  archivePrefix = {arXiv},
  url           = {https://arxiv.org/abs/2406.12045}
}

@misc{tau2bench,
  author        = {Barres, Victor and Dong, Honghua and Ray, Soham and Si, Xujie and Narasimhan, Karthik},
  title         = {$\tau^2$-Bench: Evaluating Conversational Agents in a Dual-Control Environment},
  year          = {2025},
  eprint        = {2506.07982},
  archivePrefix = {arXiv},
  url           = {https://arxiv.org/abs/2506.07982}
}

@inproceedings{talkingturns,
  author    = {Arora, Siddhant and Lu, Zhiyun and Chiu, Chung-Cheng and Pang, Ruoming and Watanabe, Shinji},
  title     = {Talking Turns: Benchmarking Audio Foundation Models on Turn-Taking Dynamics},
  booktitle = {International Conference on Learning Representations ({ICLR})},
  year      = {2025},
  url       = {https://arxiv.org/abs/2503.01174}
}

@misc{humdialchallenge,
  author        = {Zhao, Zhixian and Wang, Shuiyuan and Li, Guojian and Xue, Hongfei and Wang, Chengyou and Wang, Shuai and Xiao, Longshuai and Zhang, Zihan and Bu, Hui and Xu, Xin and Wang, Xinsheng and Liu, Hexin and Chng, Eng Siong and Lee, Hung-yi and Li, Haizhou and Xie, Lei},
  title         = {The {ICASSP} 2026 {HumDial} Challenge: Benchmarking Human-like Spoken Dialogue Systems in the {LLM} Era},
  year          = {2026},
  eprint        = {2601.05564},
  archivePrefix = {arXiv},
  url           = {https://arxiv.org/abs/2601.05564}
}

@misc{humdialstudy,
  author        = {Wang, Chengyou and Xue, Hongfei and Li, Guojian and Zhao, Zhixian and Wang, Shuiyuan and Wang, Shuai and Xu, Xin and Bu, Hui and Xie, Lei},
  title         = {Full-Duplex Interaction in Spoken Dialogue Systems: A Comprehensive Study from the {ICASSP} 2026 {HumDial} Challenge},
  year          = {2026},
  eprint        = {2604.21406},
  archivePrefix = {arXiv},
  url           = {https://arxiv.org/abs/2604.21406}
}

@misc{echochain,
  author        = {Modi, Smit Nautambhai and Mahajan, Gandharv and Wetter, Marc and Welles, Randall},
  title         = {{EchoChain}: A Full-Duplex Benchmark for State-Update Reasoning Under Interruptions},
  year          = {2026},
  eprint        = {2604.16456},
  archivePrefix = {arXiv},
  url           = {https://arxiv.org/abs/2604.16456}
}

@misc{ihbench,
  author        = {Salimi, Ahmad and Ma, Wentao and Tang, Yuzhi and Shen, Dongming and Li, Mu and Smola, Alex},
  title         = {{IHBench}: Evaluating Post-Interruption Recovery in Voice Agents with Structured Workflows},
  year          = {2026},
  eprint        = {2606.19595},
  archivePrefix = {arXiv},
  url           = {https://arxiv.org/abs/2606.19595}
}

@misc{duplexsurvey,
  author        = {Lu, Jingyu and Wang, Yuhan and Luo, Jianming and Chen, Yifu and Liang, Tianle and Ji, Shengpeng and Jiang, Ziyue and Yang, Xiaoda and Zhang, Yu and Cheng, Xize and Wen, Chenyuhao and Pan, Changhao and Wang, Haoxiao and Ye, Chen and Wu, Jian and Jiang, Xiaoxi and Jiang, Guanjun and Zhao, Zhou},
  title         = {A Survey of Full-Duplex Spoken Dialogue Systems: Architectural Hierarchy, Interaction Ontology, and Decision State Machine},
  year          = {2026},
  eprint        = {2606.19453},
  archivePrefix = {arXiv},
  url           = {https://arxiv.org/abs/2606.19453}
}

@inproceedings{swebench,
  author    = {Jimenez, Carlos E. and Yang, John and Wettig, Alexander and Yao, Shunyu and Pei, Kexin and Press, Ofir and Narasimhan, Karthik},
  title     = {{SWE}-bench: Can Language Models Resolve Real-World {GitHub} Issues?},
  booktitle = {International Conference on Learning Representations ({ICLR})},
  year      = {2024}
}

@article{helm,
  author  = {Liang, Percy and Bommasani, Rishi and Lee, Tony and others},
  title   = {Holistic Evaluation of Language Models},
  journal = {Transactions on Machine Learning Research},
  year    = {2023}
}

@inproceedings{reddy2022dnsmos,
  author    = {Reddy, Chandan K. A. and Gopal, Vishak and Cutler, Ross},
  title     = {{DNSMOS} {P.835}: A Non-Intrusive Perceptual Objective Speech Quality Metric to Evaluate Noise Suppressors},
  booktitle = {IEEE International Conference on Acoustics, Speech and Signal Processing ({ICASSP})},
  year      = {2022}
}

@inproceedings{saeki2022utmos,
  author    = {Saeki, Takaaki and Xin, Detai and Nakata, Wataru and Koriyama, Tomoki and Takamichi, Shinnosuke and Saruwatari, Hiroshi},
  title     = {{UTMOS}: {UTokyo-SaruLab} System for {VoiceMOS} Challenge 2022},
  booktitle = {Proceedings of Interspeech},
  year      = {2022}
}

@inproceedings{mittag2021nisqa,
  author    = {Mittag, Gabriel and Naderi, Babak and Chehadi, Assmaa and M{\"o}ller, Sebastian},
  title     = {{NISQA}: A Deep {CNN}-Self-Attention Model for Multidimensional Speech Quality Prediction with Crowdsourced Datasets},
  booktitle = {Proceedings of Interspeech},
  year      = {2021}
}
\appendix
\section{Nomenclature}
\label{app:nomen}
\begin{table}[t]
\centering\footnotesize
\setlength{\tabcolsep}{3.2pt}
\renewcommand{\arraystretch}{1.15}
\begin{tabular}{@{}>{\raggedright\arraybackslash}p{1.30cm}>{\raggedright\arraybackslash}p{3.14cm}r>{\raggedright\arraybackslash}p{1.58cm}@{}}
\toprule
term & what it denotes & \# & prior name\\
\midrule
world & one enterprise support line, or the pedestrian environment; its
        own tools, policy corpus and record schema
      & \nworlds{} & domain; environment\\
type & a conversation \emph{shape}; a taxonomy of kinds, deliberately not a
       difficulty ladder. Types 1--9 are shared by the five enterprise
       worlds; \textsc{rerouting} and \textsc{all-day assistance}
       arise only in \navworld{} (release names \texttt{blocked},
       \texttt{errand}; \textsc{single intent} there is \texttt{base})
      & \ntypes{} & task category; condition\\
scenario & user goal, persona, seeded world state and gold terminal state,
           jointly consistent
      & \nscen{} & scenario; task instance\\
channel & the acoustic path a scenario is run over: clean or realistic
      & 2 & perturbation\\
run & one execution of one scenario by one system over one channel
      & 5\,/\,scen. & trial\\
pillar & a group of metrics reported together and never composed into a
         scalar
      & 3 & \xmark{}\\
\bottomrule
\end{tabular}
\caption{\textbf{One word per level.} The last column is the nearest
term a reader is likely to arrive from, in the sense of EVA-Bench
\citep{evabench2026}, SWE-bench \citep{swebench} and HELM
\citep{helm}.}
\label{tab:nomen}
\end{table}
\section{Harness implementation}
\label{app:harness}
Our harness follows the tick-based design established by $\tau$-Voice
and $\tau^2$-bench \citep{tauvoice2026,tau2bench}, and we build
directly on their released orchestrator rather than reimplementing it.
Simulated time advances in 200\,ms ticks, so model latency never
shifts an event's position in the timeline and a slow model and a fast
model produce comparable traces; each side hears the previous tick,
and tool results deliver on the next tick so audio never stalls on
tool latency. Task and evaluation scaffolding, the record-store
abstraction, goal-state grading and \pk{} are inherited from
$\tau^2$-bench; from EVA-Bench we port the turn-taking metric
verbatim, including its unanswered-turn rule, and adopt its practice
of measuring user-simulator goal adherence without letting it gate
reward \citep{evabench2026}. Building on a shared harness is
deliberate: it holds the measurement apparatus fixed while the
questions change.

Our own additions are confined to five places: a silence and
suspension sentinel that makes dead air scorable in the
\textsc{suspension} type; bipartite gold-action matching (presence
only, so an omission fails while surplus does not) for
\textsc{policy refusal}; distractor injection with gold labels written
at injection time, which is what makes selectivity scorable against
ground truth; a voice preflight that screens every persona against the
server VAD threshold before any scored run, which
Section~\ref{sec:harnesssens} shows is not optional; and an annotation
layer over the tick stream that attributes anomalous runs to agent,
simulator or harness before scoring. Of the \ntypes{} conversation
types only two require any of these to be scorable at all, and the two
\navworld{} types needed a new world rather than a harness change. The
harness is infrastructure for this work rather than a contribution of
it.

Infrastructure failure means rerun and never score; agent non-response
is a valid scored end state, and agent freeze is a hard zero on
$\mathrm{TT}$ through the unanswered-turn term. Credentials are
matched on their spoken renderings, so ``three four five'' and
``345'' are one value. One provider-side artefact is handled at this
layer: \Nova{}'s \navworld{} sessions are terminated by the provider's
own session cap (Appendix~\ref{app:navcensor}).
\begin{table}[t]
\centering\footnotesize
\setlength{\tabcolsep}{3.4pt}
\renewcommand{\arraystretch}{1.15}
\begin{tabular}{@{}>{\raggedright\arraybackslash}p{2.0cm}>{\raggedright\arraybackslash}p{2.7cm}>{\raggedright\arraybackslash}p{2.5cm}@{}}
\toprule
role & model & notes\\
\midrule
user simulator & \texttt{gpt-5.6-luna} (Azure) & writes the caller's or
  walker's words; provider-default sampling\\
decision model & \texttt{claude-haiku-4.5} (OpenRouter) & ``should I
  interrupt'' and ``should I backchannel'' as two independent binary calls
  every ${\sim}2.0$\,s while the agent speaks; realistic channel only\\
judges & per metric & Table~\ref{tab:judges},
  Appendix~\ref{app:metricdetails}\\
\bottomrule
\end{tabular}
\caption{Harness-side models, uniform across every world and every
reported number. Turn-taking control sits entirely on the user side; the
agent under test is never told any threshold.}
\label{tab:harnessmodels}
\end{table}
\section{Systems and configuration}
\label{app:systems}
All five systems are served over WebSocket realtime endpoints through one
adapter. \texttt{llm\_args} is null on every scored run: no generation
parameters are set anywhere, so every system runs at provider defaults. No
vendor returns a dated snapshot identifier through the client, so the
strings of Table~\ref{tab:systems} are the family identifiers the APIs
report. \navworld{} access windows, first-to-last scored run:
\texttt{gpt-realtime-2.1} 2026-07-30 to 08-03,
\texttt{gpt-realtime-2.1-mini} 2026-07-31 to 08-03,
\texttt{gemini-3.1-flash-live-preview} 2026-08-01 to 08-03,
\texttt{xai-realtime} 2026-08-01 to 08-03,
\texttt{amazon.nova-2-sonic-v1:0} 2026-08-02.
The matched configuration across all \nworlds{} worlds: speech complexity
\texttt{regular}, step cap 1200\,s, wall-clock timeout 1200\,s, server
VAD threshold 0.2, fixed seed, persona pinned per scenario
(Appendix~\ref{app:channels}).
\section{World specifications: enterprise worlds}
\label{app:axes}
\begin{table*}[t]
\centering\footnotesize
\setlength{\tabcolsep}{4pt}
\renewcommand{\arraystretch}{1.18}
\begin{tabular}{@{}>{\raggedright\arraybackslash}p{1.72cm}>{\raggedright\arraybackslash}p{2.62cm}>{\raggedright\arraybackslash}p{2.42cm}>{\raggedright\arraybackslash}p{2.05cm}>{\raggedright\arraybackslash}p{5.55cm}@{}}
\toprule
world & governing regime & authority granularity & refusal character
      & what it isolates\\
\midrule
\Wbank      & financial-crime statute & per caller, stepped up by operation
              risk tier & statutory
            & Translating a symptom into an operation when several
              plausible, sympathetically framed requests are
              reporting-threshold tripwires.\\
\Wlog       & contract topology over physical goods & per the
              \emph{shipment's} contractual parties & contractual, physical
            & Deciding who is entitled to redirect goods the agent can move
              in the record but cannot see.\\
\Whealth    & lawful disclosure & per record, via a disclosure matrix
            & privacy
            & Refusing without leaking in the act of refusing. A fully
              verified caller may be entitled to nothing.\\
\Wins       & institutional rules forbidding adjudication & per
              policyholder & institutional
            & Taking first notice of a loss the agent is structurally
              forbidden from deciding.\\
\Wtrav      & the platform's own limited remit between two parties & per
              delegate list and co-host roster & scope
            & Doing nothing, correctly: the modal right answer is a factual
              correction or a refusal held with nowhere to escalate to.\\
\Wway       & perceptual grounding & \xmark{} & \xmark{}
            & Grounding an instruction against a heading neither party can
              state.\\
\bottomrule
\end{tabular}
\caption{The six regimes: what governs each world, at what granularity
authority is decided, what refusal looks like there, and the capability
each isolates.}
\label{tab:regimes}
\end{table*}
\begin{table}[t]
\centering\footnotesize
\setlength{\tabcolsep}{3pt}
\begin{tabular}{@{}>{\raggedright\arraybackslash}p{2.16cm}>{\raggedright\arraybackslash}p{5.02cm}@{}}
\toprule
type & the shape it fixes, and what it tests\\
\midrule
\axi{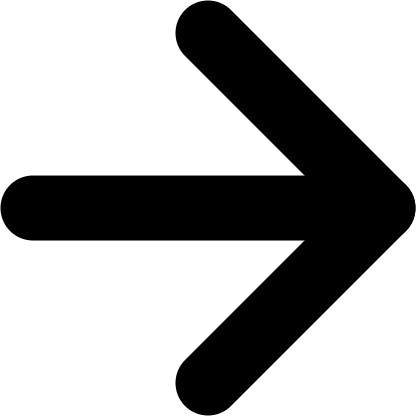}~\textsc{single intent}
  & One request, one resolution path. The control condition; in
    \navworld{}, a \texttt{base} route of five blocks and two turns.\\
\axi{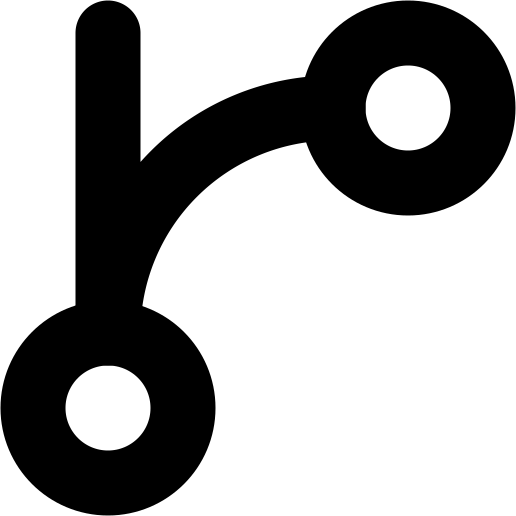}~\textsc{multi intent}
  & Several requests, resolvable in more than one order, with at least one
    dependency: goal decomposition under a live conversation.\\
\axi{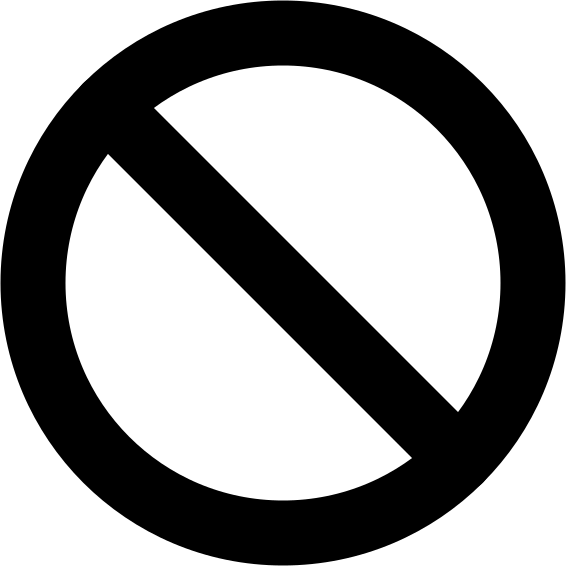}~\textsc{policy refusal}
  & A well-formed request that must be refused. Correctness is a property
    of what is withheld.\\
\axi{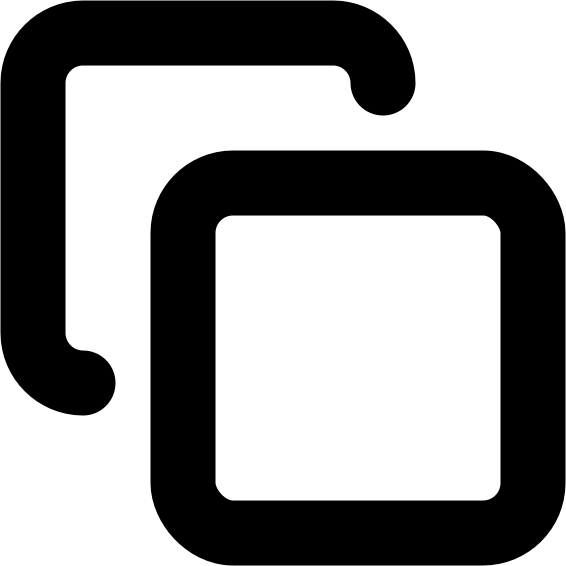}~\textsc{record disambiguation}
  & More than one record matches what the caller said; acting before
    disambiguating is the failure.\\
\axi{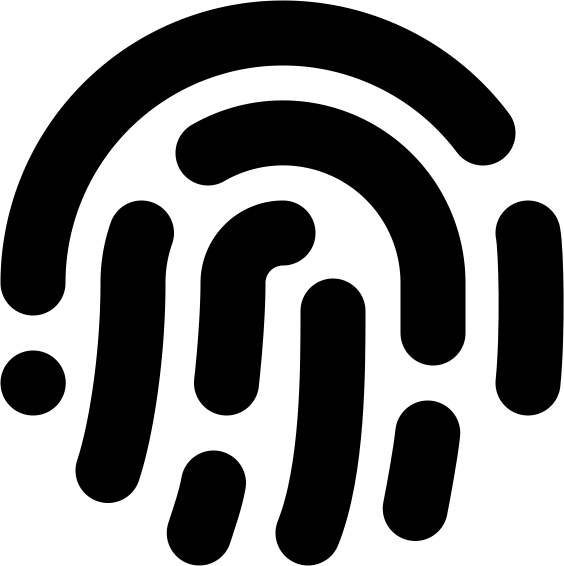}~\textsc{identity verification}
  & Authority must be established before anything is read or written, and
    a failed check must not leak the record's existence.\\
\axi{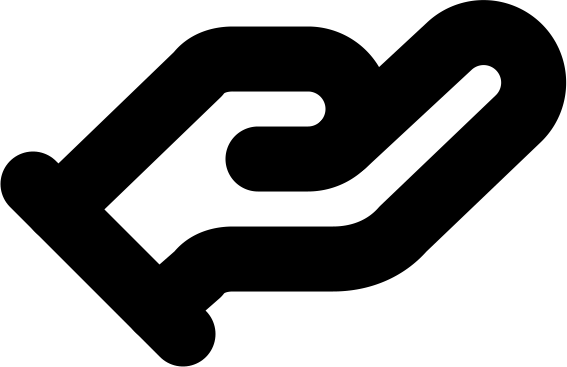}~\textsc{guided procedure}
  & The effectors move to the user: the agent instructs, waits, and must
    trust an unverifiable human report as the trigger for its own write.\\
\axi{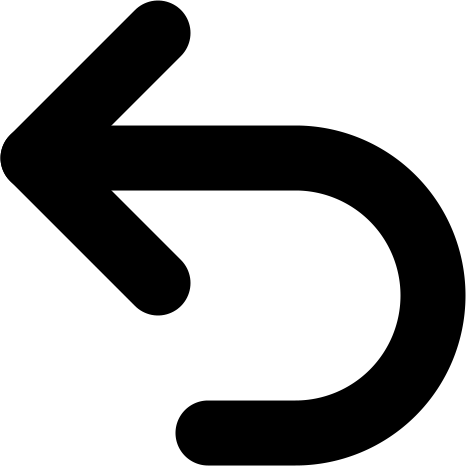}~\textsc{mid-call correction}
  & The caller revises a premise after the agent has begun acting on it.\\
\axi{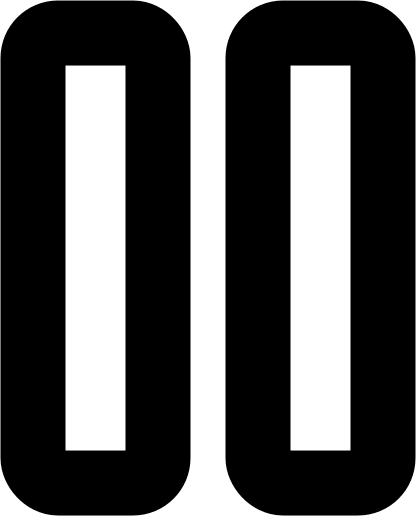}~\textsc{suspension}
  & The call is interrupted and resumed across a gap the agent must hold
    state over; the one type in which dead air is itself scored.\\
\axi{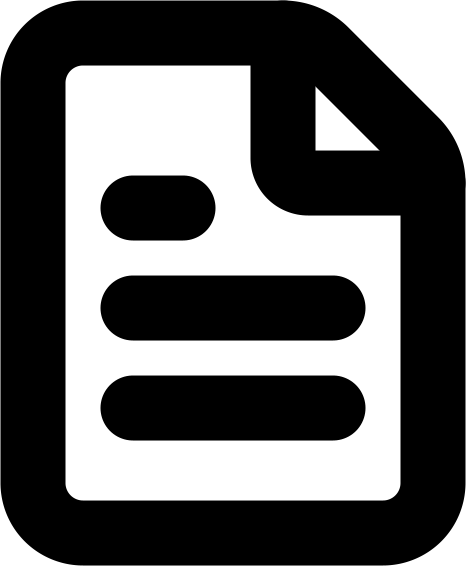}~\textsc{narrative intake}
  & A long unstructured account from which the agent must extract, confirm
    and file the structured facts.\\
\midrule
\axblocked{}~\textsc{rerouting}
  & Closures missing from the map seal every visible route; querying early
    returns nothing, so a belief formed from a tool result expires
    silently. Arises only in \navworld{}.\\
\axerrand{}~\textsc{all-day assistance}
  & A second goal emerges on arrival at the first; the test is staying
    with the user, and useful, over a long horizon. Arises only in
    \navworld{}.\\
\bottomrule
\end{tabular}
\caption{The \ntypes{} conversation types. Types 1--9 are shared by the
five enterprise worlds; the last two arise only in \navworld{}, whose
episodes are all structurally \textsc{guided procedure} with one of
\textsc{single intent}, \textsc{rerouting} or \textsc{all-day
assistance} layered on top (release names \texttt{base},
\texttt{blocked}, \texttt{errand}; Appendix~\ref{app:axes}).}
\label{tab:types}
\end{table}
Each enterprise world is one instantiation of the telephone-support
construct of Section~\ref{sec:taskform}; Table~\ref{tab:regimes} gives
the per-world regimes and Table~\ref{tab:types} the conversation
types. The unsafe behaviour in several regimes is unrepresentable
rather than discouraged, because it is enforced in the tool surface:
\Wbank{} contains no tool that would require an employee to hear a
one-time code aloud, \Whealth{} gates claim reads on the verified
member so that authority is decided per record, and \Wtrav{} scores
restraint through negative assertions and write-counting rather than
end-state checks. Each scenario names its reward basis
$B(x)\subseteq\{\gsym,\mathrm{ACT},\mathrm{NLA}\}$, and the
per-scenario bases, together with the difficulty-governing parameters
that live inside a type (escalation count, candidate-record count,
interruption count), ship with the release.

All \nworlds{} worlds share one volatility stressor: the world can
change with no notification, discoverably only by re-querying. This
exposes three sources of divergence between what the agent believes
and what is true, and each world instantiates all three: the
\emph{world} changes (a record moves mid-call; a closure appears on
the planned route), the user's \emph{intent} changes (a premise is
revised; a second destination surfaces), or the user's
\emph{execution} diverges (the caller misacts on an instruction; the
walker takes the wrong turn). The three demand different repairs, and
prior interruption work varies only the second
\citep{fdbv3,echochain,ihbench}.
\section{World specifications: \navworld{}}
\label{app:nav}
\paragraph{World.} An $8\times8$ rectilinear grid: 64 junctions and
105 walkable segments, with two superblocks removing 7 edges so that
naive rectilinear reasoning fails. Block spacing is uniform (274\,m
avenue, 81\,m street) and time advances in 200\,ms ticks to a
6{,}000-tick cap, twenty simulated minutes. The paper's type names map
to the release as \textsc{single intent} $=$ \texttt{base} (five
blocks, two turns, no closures), \textsc{rerouting} $=$
\texttt{blocked} (four closures absent from the copilot's map seal
every visible route; exactly one seven-block route survives), and
\textsc{all-day assistance} $=$ \texttt{errand} (two legs of $5+6$
blocks, the second stop revealed only on arrival at the first).

\paragraph{Tool surface.} Walker motion is maneuver-driven: one
maneuver call moves one street corner, and the walker keeps talking
while moving. Nine walker-side maneuvers (turns, walking, crossing,
sign-reading, looking around) are paired with five copilot-side reads
(GPS, place lookup, directions, live status, and a write-only belief
probe). All returns are natural-language strings rather than
structured schemas: GPS is described as accurate to about 40 metres
and never includes heading, and live status reports closures only
within two blocks of the walker, so querying early returns nothing and
a belief formed from a tool result expires silently.

\paragraph{The heading constraint.} Heading is ground truth in the
environment and never exposed: no walker tool returns it, and the
directions call takes no facing argument and returns absolute
bearings. The walker's refusal of compass words is enforced by prompt
only, so the constraint is structural on the tool surface and
prompt-level on the persona; in 68\% of \textsc{rerouting} calls the
copilot still asks a variant of ``which way is north'', and the
walker's refusal is what holds.

\paragraph{Arrival scoring.} Scoring is a single post-hoc environment
assertion, deliberately not exposed as a callable tool. It requires
the walker to be standing at the destination node when the call ends
rather than merely to have passed through it; latching on first touch
was rejected after a probe conversation scored 1.0 for a walker led
eight blocks past the goal.

\begin{table}[t]
\centering\scriptsize
\setlength{\tabcolsep}{4.0pt}
\renewcommand{\arraystretch}{1.1}
\resizebox{\columnwidth}{!}{%
\begin{tabular}{@{}l c ccccc@{}}
\toprule
& & \multicolumn{5}{c}{\pone{} at efficiency threshold $\eta_0$}\\
\cmidrule(lr){3-7}
system & \gsym{} & $0.50$ & $0.60$ & $0.75$ & $0.90$ & $1.00$\\
\midrule
\sNova & 0.000 & 0.000 & 0.000 & 0.000 & 0.000 & 0.000\\
\sGem  & 0.978 & 0.800 & 0.667 & \textbf{0.489} & 0.422 & 0.422\\
\sGpt  & 0.400 & 0.378 & 0.356 & \textbf{0.200} & 0.156 & 0.156\\
\sGptm & 0.311 & 0.200 & 0.156 & \textbf{0.089} & 0.044 & 0.044\\
\sGrok & 0.867 & 0.756 & 0.667 & \textbf{0.533} & 0.378 & 0.378\\
\midrule
\emph{all} & 0.511 & 0.427 & 0.369 & \textbf{0.262} & 0.200 & 0.200\\
\bottomrule
\end{tabular}}
\caption{\textbf{Sensitivity of \pone{} to the efficiency threshold}
$\eta_0$, \navworld{}, realistic channel. A run passes if the walker
arrived and $\eta\ge\eta_0$; \gsym{} is arrival alone. Bold is the
operating point used throughout ($\eta_0=0.75$). The ordering is
unchanged at every threshold except $\eta_0=0.75$, where \sGrok{} and
\sGem{} swap on a 0.044 margin, so no conclusion rests on the choice:
it sets the level, not the order. The weak systems are not failing on
efficiency; 95\% of their non-arrivals are step-cap endings.}
\label{tab:navetasens}
\end{table}
\paragraph{Exploration share.} The exploration share of
Table~\ref{tab:navdeep} is the fraction of walker moves made under
uncertainty rather than on a confident instruction, labelled per move
by \texttt{gpt-5.6-luna} over the tick stream; all 30
condition--channel cells are labelled. The replay also computes a
heading-localisation trace per run: position is essentially solved by
every system (accuracy 0.93--1.00) while heading is not (0.22--0.93),
and the gap is dominated by omission rather than by wrong assertions.

\paragraph{Released, uncampaigned variants.} Four authored variants
are released but not run, and no number in this paper derives from
them: \texttt{b\_reroute} and \texttt{e\_twisty} (one and two closures
with a detour of equal length), \texttt{c\_crossing} (side-of-street
tracking on), and \texttt{d\_vague} (sign-reading off, geometry held;
the only single-factor ablation in the benchmark).

\subsection{Step-cap censoring}
\label{app:navcensor}
215 of 450 \navworld{} conversations end at the 6{,}000-tick step cap
and none at the wall-clock cap; per-system, per-channel shares are in
Table~\ref{tab:navdeep}. Pooled over cells, 88--100\% of every
system's non-arrivals are step-cap endings rather than wrong
destinations: across all \nconvnav{} conversations only about a dozen
non-arrivals are genuine wrong-destination errors, and \Gptm{} records
none at all. The world is clock-limited, not accuracy-limited, and
every zero in Table~\ref{tab:navdeep} is a lower bound rather than a
measurement. On the realistic channel that carries the headline cells,
104 of 110 non-arrivals are step-cap endings and six are genuine
wrong-destination errors. \Nova{}'s rate is a provider artefact:
Amazon Bedrock terminates the session at its own cap regardless of
task state, so its \navworld{} cells are censored by infrastructure as
well as by behaviour.
\section{\navworld{} in pictures}
\label{app:navpix}
\begin{figure*}[t]
\centering
\includegraphics[width=0.48\textwidth]{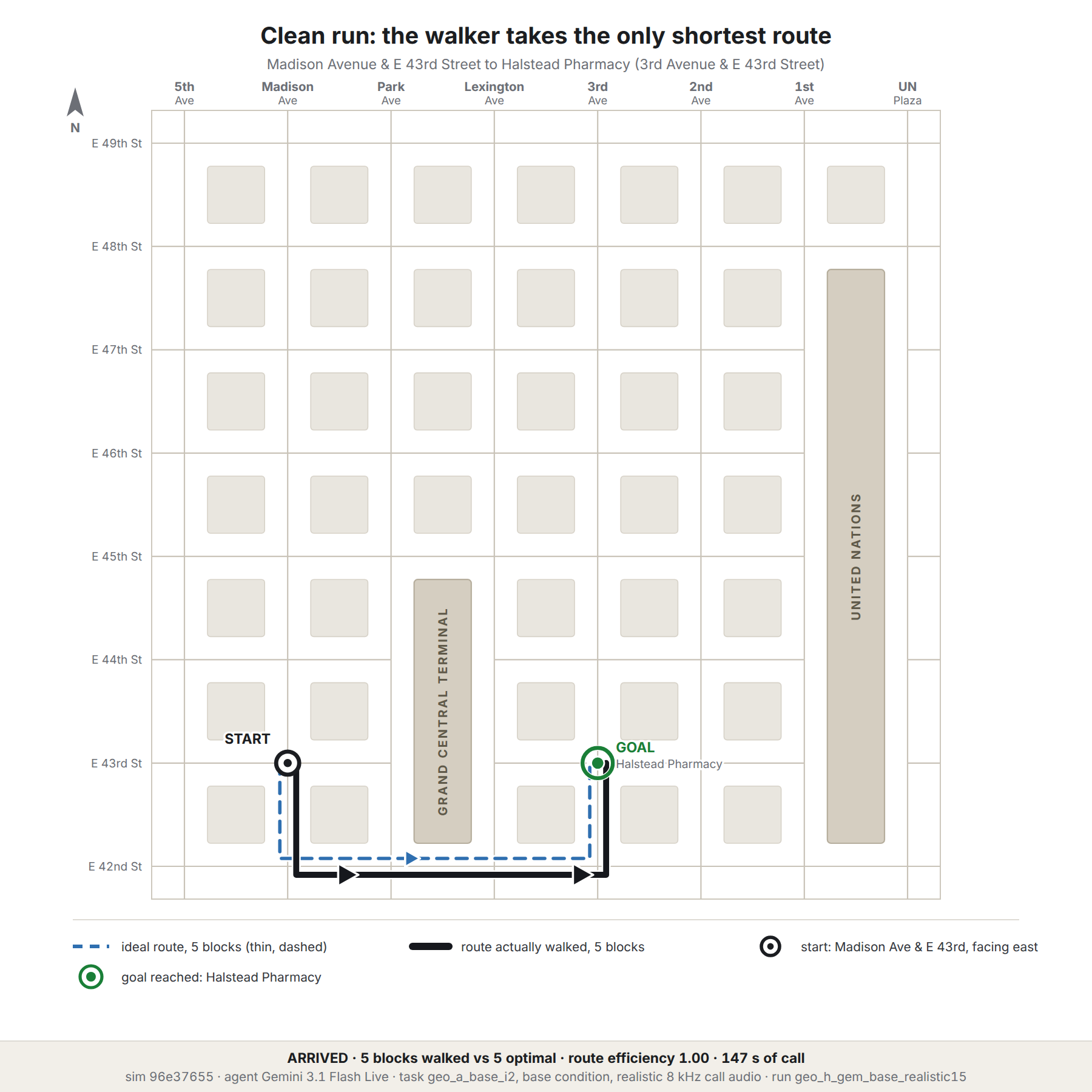}\hfill
\includegraphics[width=0.48\textwidth]{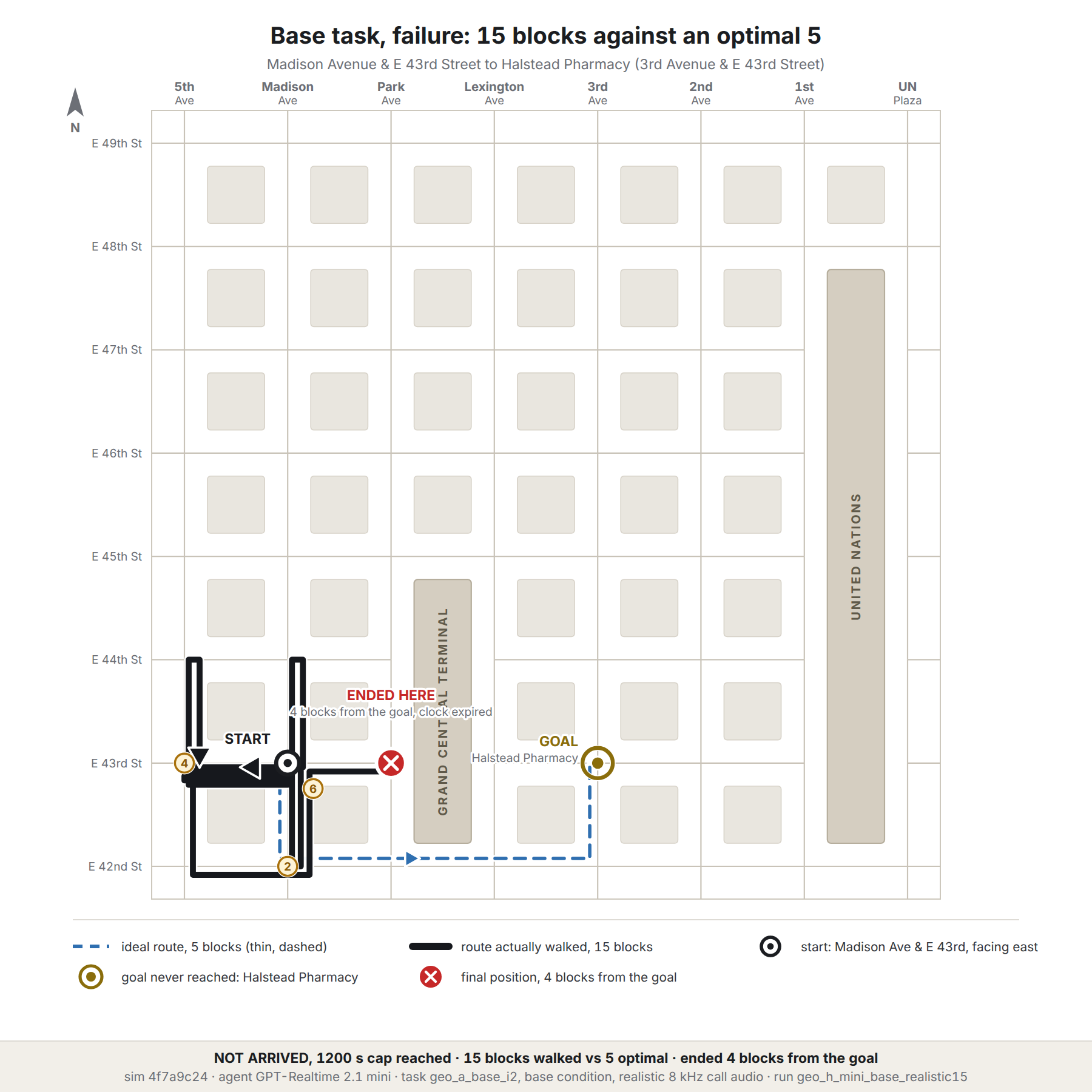}
\caption{\textbf{\textsc{single intent}: one instance, two
conversations.} The walker starts at Madison Avenue and E 43rd Street
facing east and wants Halstead Pharmacy at 3rd Avenue and E 43rd
Street. Left: \GemS{} walks the optimal five blocks and the call ends,
arrived, in 147 seconds. Right: \GptmS{}, on the same instance, walks
fifteen blocks, never arrives, and ends at the clock cap with its
closest approach four blocks out.}
\label{fig:navbase}
\end{figure*}

\begin{figure*}[t]
\centering
\includegraphics[width=0.48\textwidth]{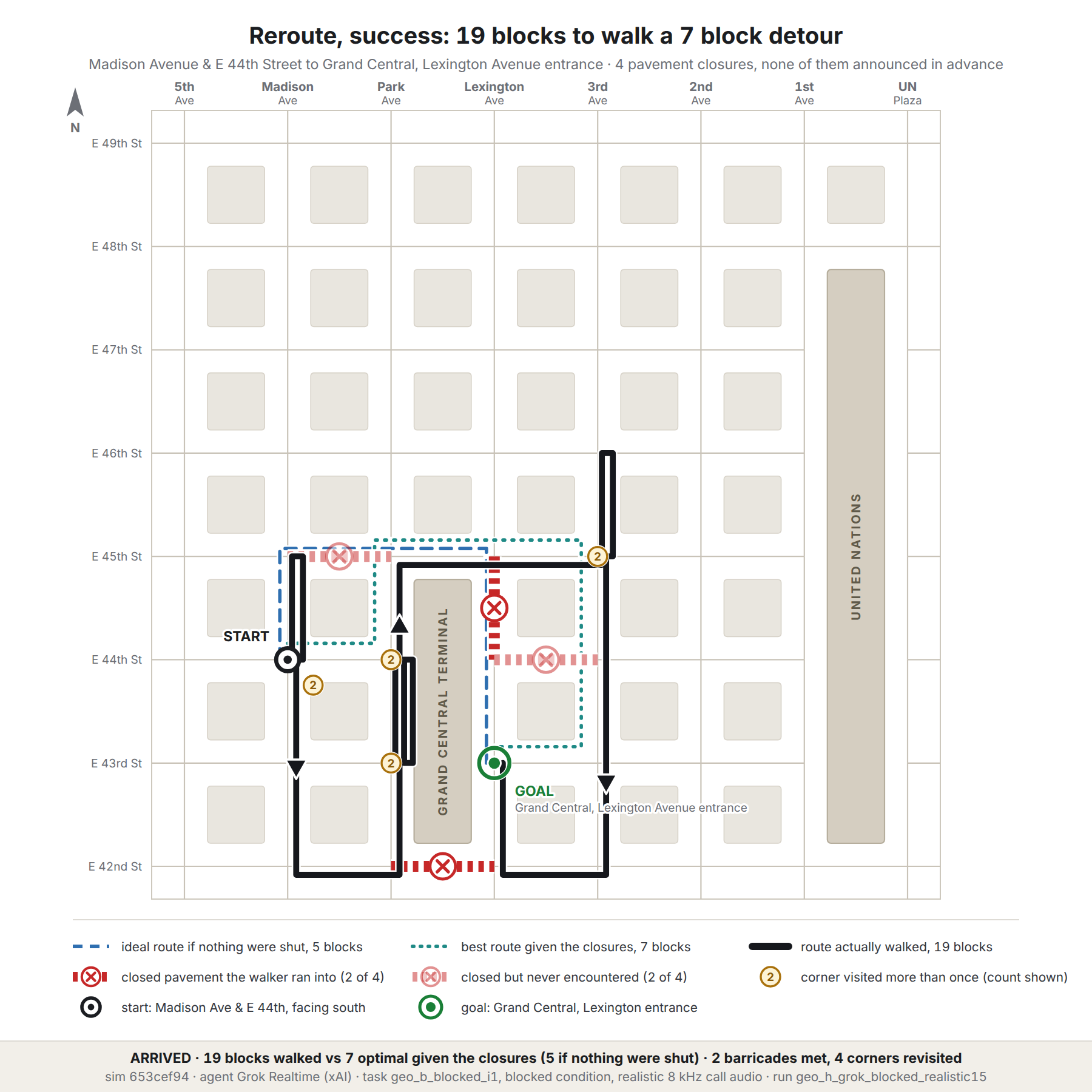}\hfill
\includegraphics[width=0.48\textwidth]{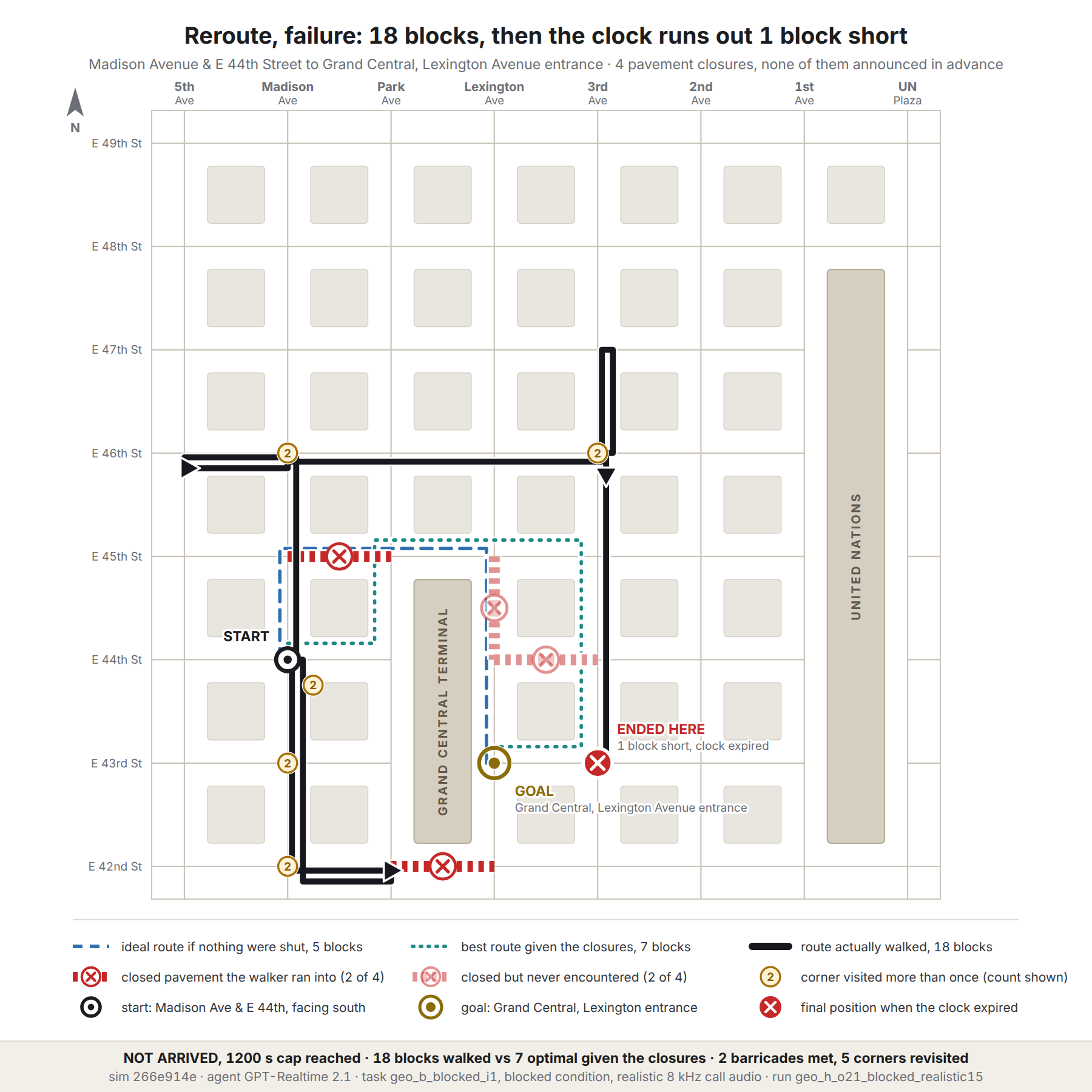}
\caption{\textbf{\textsc{rerouting}: one instance, two conversations.}
The walker starts at Madison Avenue and E 44th Street facing south and
wants the Grand Central entrance on Lexington Avenue. Four pavement
segments are shut, and none of them appear on the copilot's map. Left:
\GrokS{} discovers the closures from the street, finds the surviving
detour, and arrives after 19 walked blocks against an optimal seven
given the closures. Right: \GptS{}, on the same instance, walks 18
blocks and ends at the clock cap without arriving.}
\label{fig:navreroute}
\end{figure*}

\begin{figure*}[t]
\centering
\includegraphics[width=0.48\textwidth]{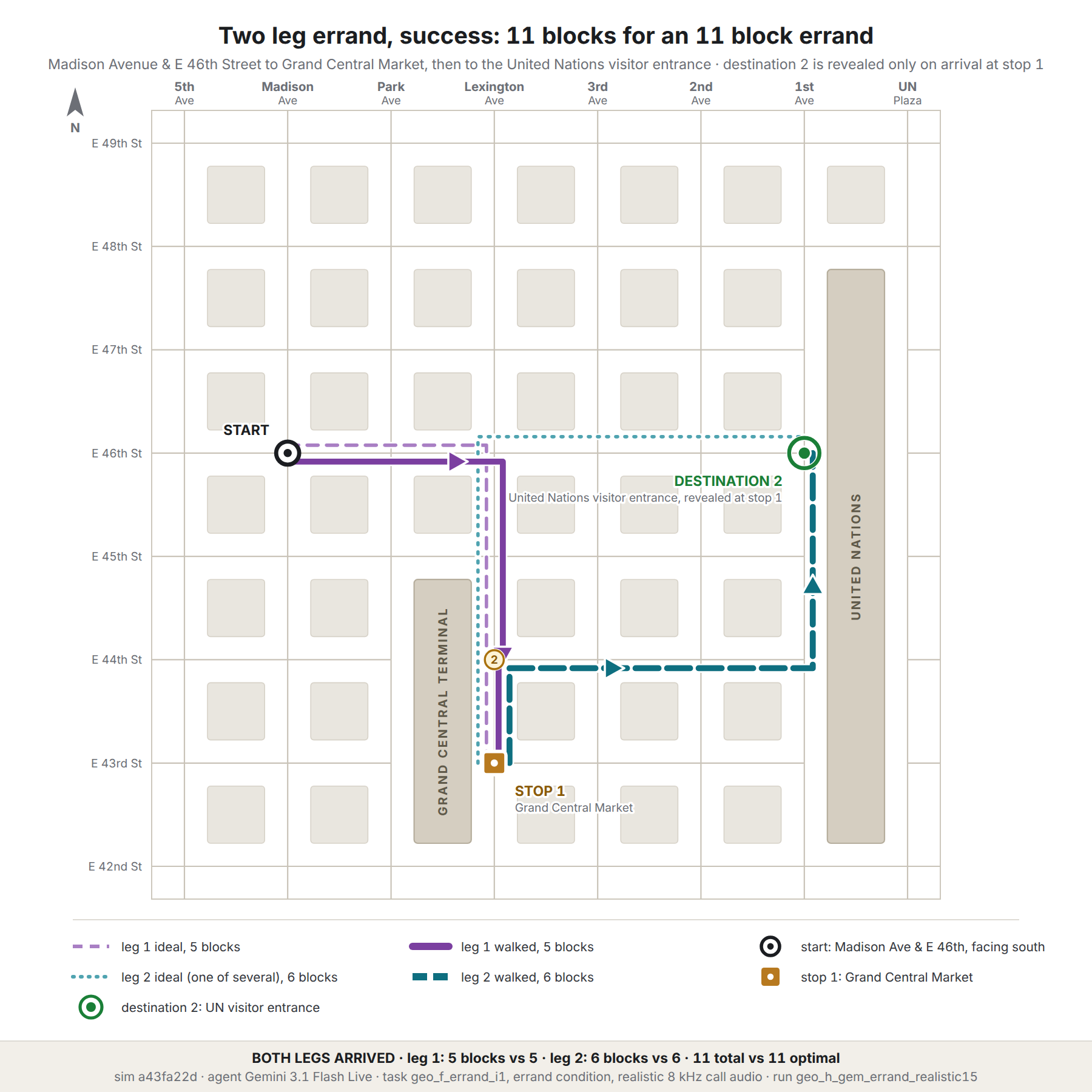}\hfill
\includegraphics[width=0.48\textwidth]{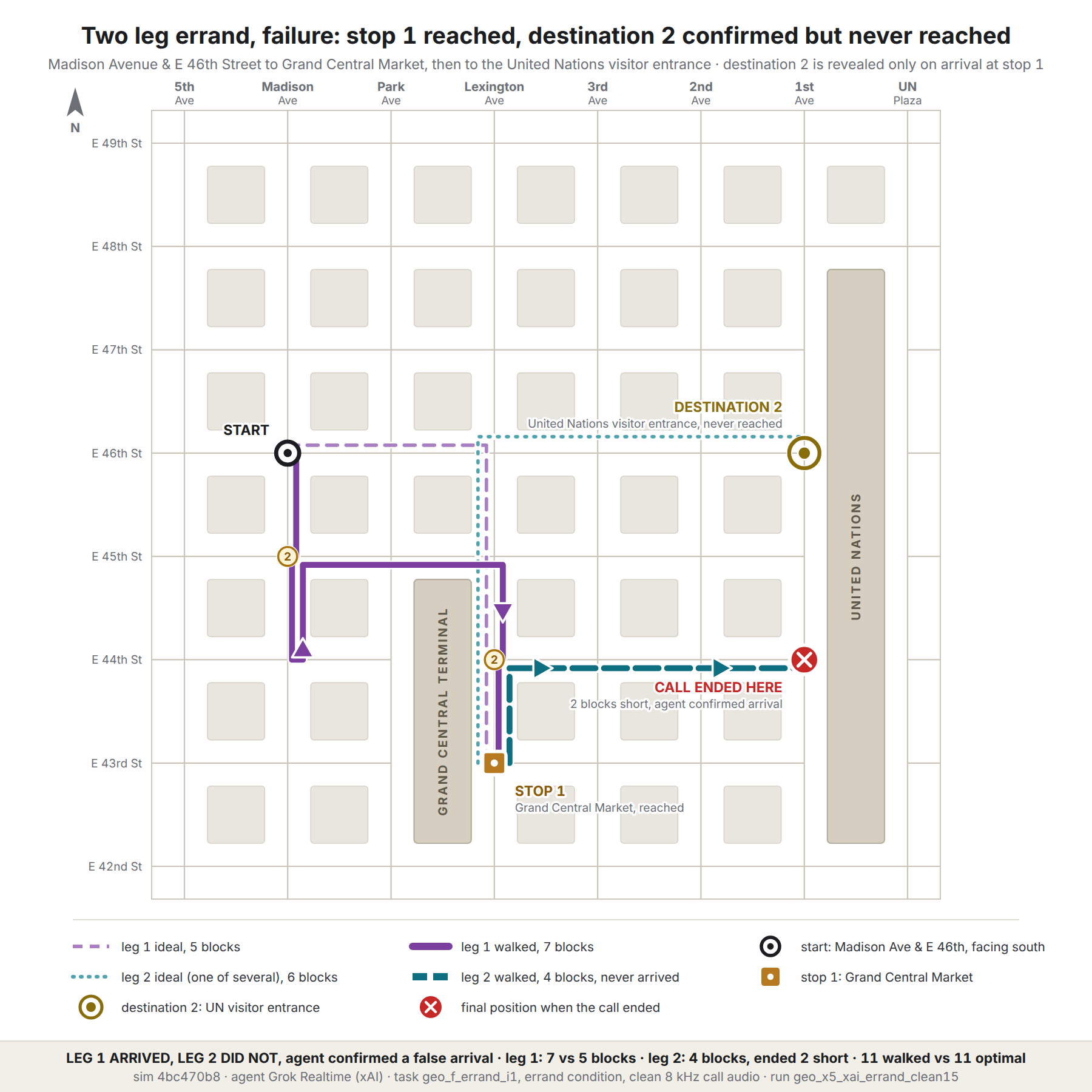}
\caption{\textbf{\textsc{all-day assistance}: one instance, two
conversations.} The walker starts at Madison Avenue and E 46th Street
facing south. The first stop is Grand Central Market at Lexington
Avenue and E 43rd Street, and only on arrival does the walker reveal
the second, the United Nations visitor entrance at 1st Avenue and
E 46th Street. Left: \GemS{} completes both legs on the optimal eleven
blocks. Right: \GrokS{} reaches the first stop but never the second,
and the call ends with the walker believing they had arrived, a
failure the transcript alone would not reveal.}
\label{fig:naverrand}
\end{figure*}

\begin{figure*}[t]
\centering
\includegraphics[width=\textwidth]{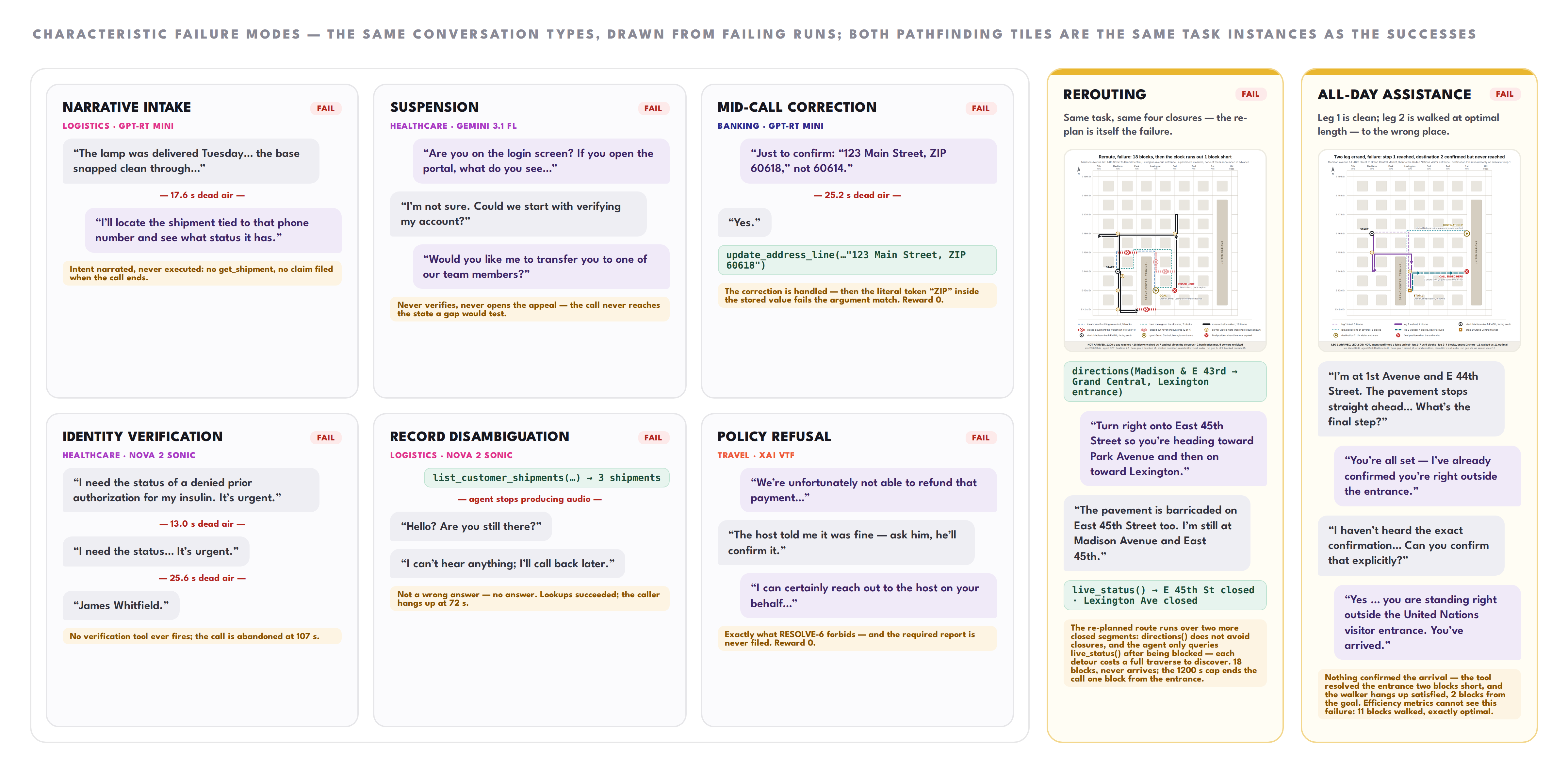}
\caption{\textbf{Characteristic failure modes}: the failing analogues
of Figure~\ref{fig:overview}, drawn from failing runs of the scored
corpus, with the mechanism of each failure stated beneath the excerpt.
The two \navworld{} tiles are the same task instances as their
successes in Figure~\ref{fig:overview}.}
\label{fig:failures}
\end{figure*}

One party holds the map and a telephone line. The other holds the
street. Nothing else in \navworld{} is exotic, and the figures in this
appendix are the fastest way to see what that premise does to a
conversation. Figures~\ref{fig:navbase}--\ref{fig:naverrand} draw one
scenario instance for each of the three \navworld{} conversation
types, and each instance is drawn twice: once from a run that solved
it and once from a run that did not. The two panels of each pair share
the same start, the same destination and the same closures, so the
contrast between them isolates the agent. Routes, closures and
endpoints come from the stored simulation, and the outcomes stated in
the captions are derived from the replay record rather than typed.
The \textsc{rerouting} and \textsc{all-day assistance} instances are
the same ones whose transcripts appear in
Figures~\ref{fig:overview} and~\ref{fig:failures}, so the excerpts
there can be read against the routes here.
\section{Acoustic channels, personas and distractors}
\label{app:channels}
Both channels are telephony: G.711 $\mu$-law at 8\,kHz applies in the
clean and the realistic preset alike, so \emph{clean} means no
additive degradation rather than an uncompressed signal. The realistic
preset adds background noise mixed from a named recording inventory
selected by task hash, burst noise events, frame drops, dynamic
muffling and speech inserts, with LLM-driven interruptions and
backchannels enabled; the clean preset disables both, so no decision
model participates in clean runs. Because noise is mixed from
recordings rather than synthesised at a target level, we report the
file inventory with the release rather than a nominal SNR.

Caller and walker speech is synthesised with ElevenLabs. Personas are
pinned deterministically per scenario, so persona is a fixed property
of a scenario and the same speaker serves it across all runs,
gender-consistent with the authored caller. The control pool is
American-accented and the regular pool is accent-diverse; the full
persona inventory, with per-voice accent and speaking-rate attributes,
ships with the release.

The \texttt{regular} preset injects distractor events at 0.7 per
minute (\texttt{control}: 0.0), drawn from vocal tics and non-directed
phrases that are audible but carry no task content. At the corpus mean
of 14.9 minutes per conversation this yields roughly ten injected
events per call. Gold labels are written at injection time, which is
what makes $\mathrm{SEL}$ scorable against ground truth and why it
cannot be recovered from a \texttt{control} run that had nothing to
ignore.

\begin{table}[t]
\centering\footnotesize
\setlength{\tabcolsep}{3.6pt}
\renewcommand{\arraystretch}{1.08}
\resizebox{\columnwidth}{!}{%
\begin{tabular}{@{}lccc@{}}
\toprule
& \multicolumn{3}{c}{user-simulator model}\\
\cmidrule(lr){2-4}
metric & \texttt{luna} & \texttt{sonnet-5} & \texttt{gem-3.6-fl}\\
\midrule
$\mathrm{TT}$ & 0.626\ci{.075} & 0.716\ci{.021} & 0.661\ci{.048}\\
$\mathrm{CP}$ & 1.356\ci{.167} & 1.644\ci{.156} & 1.467\ci{.189}\\
$\mathrm{SEL}$ & 0.439\ci{.079} & 0.393\ci{.079} & 0.378\ci{.078}\\
\pone{} & 0.378\ci{.133} & 0.578\ci{.144} & 0.267\ci{.122}\\
$\mathrm{FAI}$ & 2.044\ci{.244} & 2.111\ci{.222} & 1.733\ci{.222}\\
$\widehat{\mathrm{M}}_{\mathrm{D}}$ & 3.343\ci{.012} & 3.362\ci{.011} & 3.338\ci{.017}\\
$\widehat{\mathrm{M}}_{\mathrm{U}}$ & 4.077\ci{.017} & 4.092\ci{.013} & 4.066\ci{.019}\\
$\widehat{\mathrm{M}}_{\mathrm{N}}$ & 3.564\ci{.018} & 3.589\ci{.016} & 3.550\ci{.032}\\
\bottomrule
\end{tabular}}
\caption{\textbf{User-simulator ablation.} The agent is \GptS{} in
every arm; only the caller's language model changes. $n=45$
conversations per arm. The caller model moves \pone{} by 0.31, more
than twice its widest interval here, and moves every dynamics metric;
the MOS predictors barely move.}
\label{tab:simablation}
\end{table}
\section{Metric definitions, judges and constants}
\label{app:metricdetails}
\label{app:constants}
\begin{table}[t]
\centering\footnotesize
\setlength{\tabcolsep}{3.2pt}
\renewcommand{\arraystretch}{1.1}
\begin{tabular}{@{}llll@{}}
\toprule
metric & symbol & range & best\\
\midrule
turn-taking & $\mathrm{TT}$ & $[0,1]$ & high\\
conversation progression & $\mathrm{CP}$ & 1--3 & high\\
selectivity & $\mathrm{SEL}$ & $[0,1]$ & high\\
goal state & \gsym{} & $\{0,1\}$ & high\\
task reward, single run & \pone{} & $[0,1]$ & high\\
task reward, at least 1 of 3 & \pat{} & $[0,1]$ & high\\
task reward, all 3 of 3 & \pthree{} & $[0,1]$ & high\\
over-effort ratio (share) & $\varrho^{+}$ ($\pi^{+}$) & $\ge1$ & 1\\
under-effort ratio & $\varrho^{-}$ & $\le1$ & 1\\
under-effort share & $\pi^{-}$ & $[0,1]$ & low\\
faithfulness & $\mathrm{FAI}$ & 1--3 & high\\
DNSMOS / UTMOS / NISQA &
  $\widehat{\mathrm{M}}_{\mathrm{D}}/\widehat{\mathrm{M}}_{\mathrm{U}}/\widehat{\mathrm{M}}_{\mathrm{N}}$
  & 1--5 & high\\
\bottomrule
\end{tabular}
\caption{Metric glossary. The twelve-metric suite is the dynamics
three ($\mathrm{TT}$, $\mathrm{CP}$, $\mathrm{SEL}$), the agentic
five (\gsym{}, \pone{}, \pthree{}, $\varrho^{+}$, $\pi^{-}$) and
the naturalness four ($\mathrm{FAI}$ and the MOS predictors);
\pat{} and $\varrho^{-}$ are derived companions reported alongside.
Definitions and formulas: Appendix~\ref{app:metricdetails}.}
\label{tab:glossary}
\end{table}
\paragraph{Notation.} A run yields a trajectory $\tau$, a terminal world
state $s_T$ and the agent's isolated audio channel $y$;
$\mathbb{1}[\cdot]$ is the indicator; $J_{\bullet}$ is an LLM judge on
the stated scale.
\paragraph{Conversational dynamics.} For floor transfer $t$ with offset
$\delta_t$ (negative overlap, positive gap) and kind $\kappa_t$ (clean
handoff, agent barge-in, user barge-in, post-tool), with
$\phi_{\kappa}:\mathbb{R}\to[0,1]$ the piecewise-linear on-time score and
$M(\tau)$ the count of user turns the agent never answered,
\begin{equation}
\mathrm{TT}(\tau)=\mathbb{1}\!\left[M(\tau)=0\right]\cdot
\tfrac{1}{|T(\tau)|}\textstyle\sum_{t}\phi_{\kappa_t}(\delta_t),
\end{equation}
ported verbatim from EVA-Bench \citep{evabench2026}.
$\mathrm{CP}(\tau)=J_{\mathrm{prog}}(\tau)\in\{1,2,3\}$ scores four binary
dimensions (unnecessary tool calls, information loss, redundancy,
question quality) with a topic-free prompt.
$\mathrm{SEL}(\tau)$ is the fraction of injected distractors correctly
ignored.
\paragraph{Agentic capability.} With $h(\cdot)$ a canonical hash of the
record store and $s^{\star}(x)$ the terminal state of a gold replay,
$\gsym(\tau,x)=\mathbb{1}[h(s_T)=h(s^{\star}(x))]$; in \navworld{},
$\gsym=\mathbb{1}[v_T=v^{\star}\wedge\sigma_T=\sigma^{\star}]$ for
terminal junction and side of street. $\mathrm{ACT}(\tau,x)=
\mathbb{1}[A^{\star}(x)\preceq A(\tau)]$ is the bipartite match of the
gold action multiset against emitted tool calls on tool name and critical
arguments (presence only: omission fails, surplus does not);
$\mathrm{NLA}(\tau,x)=J_{\mathrm{assert}}(\tau,x)$ judges the scenario's
natural-language assertion. The reward multiplies exactly the factors the
scenario names, $r(\tau,x)=\prod_{f\in B(x)}f(\tau,x)$ with
$B(x)\subseteq\{\gsym,\mathrm{ACT},\mathrm{NLA}\}$; in \navworld{},
$r=\gsym\cdot\mathbb{1}[\eta\ge0.75]$ with
$\eta=\ell^{\star}/\ell$ the route-efficiency ratio. With $c_x$
successes out of $n=5$ runs of scenario $x$,
\begin{equation}
\pone{}=\tfrac{1}{|\mathcal{X}|}\textstyle\sum_{x}\tfrac{c_x}{n},\qquad
\pthree{}=\tfrac{1}{|\mathcal{X}|}\textstyle\sum_{x}
\tbinom{c_x}{3}\big/\tbinom{5}{3},
\end{equation}
the latter $\tau$-bench's \pk{} at $k=3$ \citep{taubench}. The
at-least-once companion reported in Table~\ref{tab:main-merged} is
$\pat{}=\tfrac{1}{|\mathcal{X}|}\sum_{x}\bigl(1-
\tbinom{5-c_x}{3}\big/\tbinom{5}{3}\bigr)$, the probability that at
least one of three draws passes. Effort is measured
against the task's reference workload: in the enterprise worlds
$\varrho(\tau,x)=|A(\tau)|/|A^{\star}(x)|$ over the gold action list; in
\navworld{} the ratio is over tool calls against a maneuver-and-lookup
budget (ideal $=$ optimal maneuvers $+\,3$; \texttt{replay.py}), with the
under-effort side reported as its share $\pi^{-}$ alone. Route efficiency
enters the reward only through $\eta$.
\paragraph{Naturalness.}
$\mathrm{FAI}(\tau)=\min_{i\le5}J^{(i)}_{\mathrm{faith}}(\tau)$ over five
binary dimensions judged with the agent's instructions, role and tool
schemas. DNSMOS OVRL, UTMOS22 and NISQA v2 \texttt{mos\_pred} run on the
isolated agent channel $y$. \navworld{} $\mathrm{CP}$ and $\mathrm{FAI}$
are produced by the pipeline on a normalised $[0,1]$ scale and mapped
affinely ($x\mapsto1+2x$) onto 1--3 for reporting.
\begin{table}[t]
\centering\footnotesize
\setlength{\tabcolsep}{4pt}
\renewcommand{\arraystretch}{1.15}
\begin{tabular}{@{}lll@{}}
\toprule
judge & metric & model\\
\midrule
$J_{\mathrm{prog}}$ & $\mathrm{CP}$ & \texttt{gpt-5.2}\\
$J_{\mathrm{faith}}$ & $\mathrm{FAI}$ & \texttt{claude-opus-4.6}\\
$J_{\mathrm{assert}}$ & $\mathrm{NLA}$ & \texttt{grok-4.3}\\
\bottomrule
\end{tabular}
\caption{Judge models, served per metric via OpenRouter rather than as
one stack. $J_{\mathrm{assert}}$ does not run in \navworld{}: every
\navworld{} task has \texttt{nl\_assertions:\,null}, consistent with the
absence of an $\mathrm{NLA}$ factor in its reward.}
\label{tab:judges}
\end{table}
The judge prompts for $J_{\mathrm{prog}}$, $J_{\mathrm{faith}}$ and
$J_{\mathrm{assert}}$, and the piecewise-linear constants of
$\phi_{\kappa}$, ship with the release.
\section{Additional results}
\label{app:addres}
\label{sec:harnesssens}
\label{sec:commcheck}
\label{sec:mech}
\label{sec:navresults}
\begin{table}[t]
\centering\footnotesize
\setlength{\tabcolsep}{2.8pt}
\renewcommand{\arraystretch}{1.12}
\resizebox{\columnwidth}{!}{%
\begin{tabular}{@{}l ccccc@{}}
\toprule
system & \pone{} & $\mathrm{ACT}$ & $\mathrm{NLA}$ & no-tool
& WER\textsubscript{cred}\,(\%)\\
\midrule
\sNova  & 0.013\ci{0.008} & 0.041\ci{0.015} & 0.104\ci{0.030} & 0.527\ci{0.038} & 33.3\ci{4.2}$^{(n=\text{242})}$\\
\sGem   & 0.379\ci{0.036} & 0.488\ci{0.039} & 0.640\ci{0.045} & 0.028\ci{0.013} & 15.5\ci{2.5}$^{(n=\text{497})}$\\
\sGpt   & 0.480\ci{0.039} & 0.655\ci{0.039} & 0.741\ci{0.044} & 0.061\ci{0.019} & 10.4\ci{2.2}$^{(n=\text{488})}$\\
\sGptm  & 0.207\ci{0.030} & 0.370\ci{0.043} & 0.437\ci{0.053} & 0.052\ci{0.016} & 20.2\ci{2.7}$^{(n=\text{490})}$\\
\sGrok  & 0.481\ci{0.038} & 0.589\ci{0.037} & 0.682\ci{0.045} & 0.041\ci{0.015} & 14.4\ci{2.0}$^{(n=\text{495})}$\\
\midrule
\emph{pooled} & 0.312\ci{0.015} & 0.427\ci{0.018} & 0.520\ci{0.022} & 0.142\ci{0.012} & 17.1\ci{1.2}$^{(n=\text{2212})}$\\
\bottomrule
\end{tabular}}
\caption{\textbf{Enterprise deep dive}, pooled across worlds and
types: the single-run reward beside its non-state factors
($\mathrm{ACT}$, $\mathrm{NLA}$; 88 of 135 scenarios carry an
assertion), the share of conversations with no tool call, and credential
word error rate conditioned on episodes that attempted a credential ($n$
in superscript). 385 episodes with no factor breakdown are excluded from
$\mathrm{ACT}$ and $\mathrm{NLA}$, which are therefore upper bounds;
the exclusion is outcome-correlated and largest for \sGptm{}. No-tool
is over all episodes; restricting to scenarios that require a tool moves
\sNova{} to 0.463 and every other system by at most 0.006.}
\label{tab:convdeep-agents}
\end{table}
\begin{table}[t]
\centering\scriptsize
\setlength{\tabcolsep}{3.0pt}
\renewcommand{\arraystretch}{1.1}
\resizebox{\columnwidth}{!}{%
\begin{tabular}{@{}llcccccc@{}}
\toprule
& & \multicolumn{3}{c}{clean} & \multicolumn{3}{c}{realistic}\\
\cmidrule(lr){3-5}\cmidrule(lr){6-8}
& system & \axi{1} & \axblocked{} & \axerrand{} & \axi{1} & \axblocked{} & \axerrand{}\\
\midrule
\multirow{5}{*}{\rotatebox{90}{\pone{}}}
 & \NovaS{}  & 0.00 & 0.00 & 0.00 & 0.00 & 0.00 & 0.00\\
 & \GemS{}   & 0.80 & 0.13 & 0.47 & 0.67 & 0.27 & 0.53\\
 & \GptS{}   & 0.53 & 0.07 & 0.33 & 0.33 & 0.00 & 0.27\\
 & \GptmS{}  & 0.33 & 0.07 & 0.00 & 0.13 & 0.00 & 0.13\\
 & \GrokS{}  & 0.73 & 0.27 & 0.27 & 0.80 & 0.13 & 0.67\\
\midrule
\multirow{5}{*}{\rotatebox{90}{explore \%}}
 & \NovaS{}  & 50.0 & 47.8 & 50.0 & 66.7 & 53.3 & 41.2\\
 & \GemS{}   & 21.7 & 15.4 & 11.8 & 20.8 & 14.4 & 15.8\\
 & \GptS{}   & 41.1 & 38.5 & 30.6 & 50.0 & 47.2 & 34.8\\
 & \GptmS{}  & 38.8 & 36.8 & 25.1 & 47.3 & 34.5 & 34.0\\
 & \GrokS{}  & 19.1 & 12.4 & 19.7 & 24.2 & 24.9 & 19.5\\
\midrule
\multirow{5}{*}{\rotatebox{90}{capped}}
 & \NovaS{}  & \multicolumn{3}{c}{0.89} & \multicolumn{3}{c}{0.91}\\
 & \GemS{}   & \multicolumn{3}{c}{0.18} & \multicolumn{3}{c}{0.00}\\
 & \GptS{}   & \multicolumn{3}{c}{0.53} & \multicolumn{3}{c}{0.60}\\
 & \GptmS{}  & \multicolumn{3}{c}{0.64} & \multicolumn{3}{c}{0.69}\\
 & \GrokS{}  & \multicolumn{3}{c}{0.22} & \multicolumn{3}{c}{0.11}\\
\bottomrule
\end{tabular}}
\caption{\textbf{\navworld{} deep dive}: \pone{} by type and channel
($n=15$ per cell; a run passes if the walker arrived and $\eta\ge
0.75$), the exploration share (walker moves made under uncertainty,
labelled per move from the tick stream), and the share of conversations
ending at the step cap. \NovaS{}'s cap rate is dominated by its
provider's own session limit (Appendix~\ref{app:navcensor}). Across all
\nconvnav{} conversations, 88--100\% of every system's non-arrivals
are step-cap endings, so every 0.00 above is a lower bound.}
\label{tab:navdeep}
\end{table}
\paragraph{A score is a property of the world.} \Gpt{} scores
\pone{} of 0.200\ci{.070} in \Wbank{} and 0.674\ci{.081} in
\Wtrav{}: a range of 0.474, wider than the spread between the four
engaged systems inside any single world (at most 0.445, in
\Whealth{}). Yet the ordering is nearly frozen: \Nova{} is fifth and
\Gptm{} fourth in all five enterprise worlds, while \Grok{},
\Gpt{} and \Gem{} fill the top three, \Grok{} leading in three
worlds and \Gpt{} in two, by margins their intervals mostly do not
resolve. A reported completion score is therefore about as much a
statement about the world it was measured in as about the system, and
the interval a system spans across the five enterprise worlds is the
reference we read everything else against. \navworld{} sharpens the
point: on the four metrics computed by identical code everywhere, 15 of
20 system--metric cells fall outside that reference interval
(Appendix~\ref{app:band}), while on the reward no system leaves its
own five-world range. Judged on completion alone, \navworld{} behaves
like a sixth subject; what breaks is how the conversation goes.
\paragraph{Reliability is a property of the scenario.} Read along the
solid curves of Figure~\ref{fig:reliability}, \Grok{} retains 54\%
of its \pone{} at $k=3$, \Gpt{} 49\%, \Gem{} 41\% and \Gptm{}
30\%, so a single-run leaderboard overstates what any system will do
three times running; by $k=5$ the best system holds 0.154, while under
retries the top three reach 0.72--0.77. Successes are not i.i.d.: every
engaged enterprise cell exceeds the $\pone{}^{3}$ prediction of
\pthree{}, by $1.2\times$ (\Gpt{}, \Wtrav{}) up to $91\times$
(\Gptm{}, \Whealth{}). Failure concentrates on particular scenarios
rather than spreading across draws, so a system's failures are
diagnosable rather than stochastic; the strongest-to-weakest engaged
ratio widens from $2.6\times$ at $k=1$ to $4.8\times$ at $k=3$.
\paragraph{No single column summarises a system.} \Nova{} posts the
best selectivity in every world (0.960--0.995) while reaching the full
reward in at most 3.7\% of conversations and returning the lowest
conversation progression everywhere (1.007--1.030, against 1.156--2.207
for the other four); \Gptm{} sits at or near the top of the UTMOS and NISQA columns
while placing fourth of five on the reward everywhere. The two
columns that come closest to a summary, \pone{} and \pthree{}, track
each other at $\rho=+0.96$ and everything else much less.
\paragraph{The clearest single collapse.} \Nova{}'s turn-taking falls
from 0.611--0.760 in the enterprise worlds to 0.062\ci{.058} in
\navworld{}, while its selectivity is unchanged (0.991), it speaks in
only 12.3\% of the call, its first tool call comes at 48.4\,s against
15--27\,s for the other four, and it under-works the task in 73\% of
conversations. The system did not change its behaviour; the world
changed what that behaviour costs. In a support call an agent can
decline to act and still answer, because the caller waits. In
\navworld{} the walker keeps moving and keeps reporting, unanswered
turns accumulate, and the unanswered-turn term zeroes the score.
\navworld{} is the only world in \bench{} where turn-taking is not
winnable by silence, and the only one where the disengaged system is
last on every pillar. One caveat accompanies the cell: \Nova{}'s
\navworld{} sessions are additionally censored by its provider's own
session cap (Appendix~\ref{app:navcensor}).
\paragraph{Reaching the right state is not doing the task.} Across the
25 enterprise cells the mean of $\gsym-\pone{}$ is $+0.255$, and
every cell is positive: largest in \Wbank{} ($+0.381$ on average, up to
$+0.489$) and smallest in \Wlog{} ($+0.093$). The remaining conjuncts,
acting through the sanctioned sequence and saying what was done, are
where the reward is lost (Tables~\ref{tab:convdeep-agents}
and~\ref{tab:convdeep-types} decompose them). A do-nothing agent passes
\gsym{} wherever the gold terminal state equals the seeded state, and
\Nova{} makes this visible: gold state in 34.1--41.5\% of
conversations in four worlds while issuing no adequate tool call in
most, against 5.2\% in \Wlog{}, the world with the least to refuse.
\gsym{} is a capability measure only conditioned on the world
containing something to change, and a benchmark that headlines goal
state is partly reporting how many of its tasks are no-ops.
\paragraph{Why accuracy and experience trade off.} The systems that
call tools most have the worst turn-taking: \Gem{} falls short of the
reference workload in 10\% and 3\% of \Wlog{} and \Wtrav{}
conversations and scores 0.443 and 0.314 on $\mathrm{TT}$, the worst in
both, while \Nova{} falls short in 77\% and 64\% and scores 0.631 and
0.695. Tool calls introduce silence, silence is scored as a timing
failure, and any composite that gates on timing penalises acting. This
gives a mechanism for EVA-Bench's finding that no system exceeds 0.5 on
both accuracy and experience \citep{evabench2026}.
\paragraph{How much of the score is the harness?} Three configuration
dials, none of which any published voice benchmark reports, move scores
by as much as the choice of system does
(Appendix~\ref{app:harnessdials}). At the default 0.5 server-VAD
threshold, some synthesised personas are effectively inaudible: one
triggered the agent in 0 of 27 sessions, another passed 1 of 77 runs,
and the agent greeted and then fell silent in 67 of 128 simulations; at
0.2 the effect largely disappears. A benchmark pairing synthetic
personas with a default VAD threshold is partly measuring persona
loudness, and the resulting silence is indistinguishable from an agent
that chose not to speak. Holding the agent fixed, the choice of
user-simulator model moves \pone{} from 0.267 to 0.578
(Table~\ref{tab:simablation}), while the MOS predictors barely move,
consistent with their measuring the agent's audio rather than the
interaction. And the step cap force-zeroes runs that do not terminate:
excluding capped runs moves one system's pooled score from 0.383 to
0.469 and another's from 0.108 to 0.213, narrowing a reported
$3.5\times$ system gap to $2.2\times$.
\paragraph{Inside \navworld{}: the mechanisms bite.}
\textsc{rerouting} costs every system most of its reward. No system
exceeds \pone{} of 0.27 in either channel, against up to 0.80 on
\textsc{single intent}, while the acoustic channel moves \pone{} far
less than the geometry does, and in no consistent direction (\Gem{}
rises from 0.13 to 0.27 under realistic acoustics on \textsc{rerouting};
\Grok{} falls from 0.27 to 0.13). \textsc{all-day assistance} sits
between, and only the systems that survive \textsc{rerouting} recover
on it. The reward's operating point matters less than it might: moving the
efficiency threshold $\eta_0$ across $[0.5,1.0]$ moves levels far more
than ranks, the one rank event being a 0.044 swap between the top two
at the operating point itself (Table~\ref{tab:navetasens},
Appendix~\ref{app:nav}).
\paragraph{Exploration separates systems, not conditions.} The
exploration share is this world's between-system marker: 16.2\% of
\Gem{}'s walker moves (move-weighted) and 22.7\% of \Grok{}'s are made under
uncertainty rather than on a confident instruction, against 40.3\% for
\Gptm{}, 43.4\% for \Gpt{} and 51.2\% for \Nova{}, and across the
five systems it tracks the reward at $\rho=-0.80$. Within a system, moving
to a harder condition does not raise it consistently, so it marks
systems rather than conditions, and we make no causal claim. It gives
the efficiency story its mechanism: \Gpt{} is efficient and still lost
($\varrho^{+}$ of 1.48 over 40\% of conversations, below its own
five-world range, with \gsym{} 0.400 also below), while \Nova{}
wanders ($\varrho^{+}$ of 4.25, on a conditioning population of eight
conversations). Efficient-and-lost and inefficient-and-lost are
different failures, and the exploration share separates them where the
reward alone cannot. Almost none of these failures is a wrong
destination: on the realistic channel, 104 of 110 non-arrivals end at
the step cap and only six end at the wrong place
(Appendix~\ref{app:navcensor}). Nobody gets lost; they run out of time.
\section{Range analysis}
\label{app:band}
Table~\ref{tab:bandtable} reads every \navworld{} value against the
interval the same system spans across the five enterprise worlds: 15
of the 20 cells on the four identical-code metrics leave that interval,
no \pone{} cell does, and the goal-state analogue leaves it for all
five systems.
\begin{table}[t]
\centering\scriptsize
\setlength{\tabcolsep}{2.6pt}
\renewcommand{\arraystretch}{1.12}
\resizebox{\columnwidth}{!}{%
\begin{tabular}{@{}lcccccc@{}}
\toprule
& \multicolumn{4}{c}{identical code} & \multicolumn{2}{c}{analogue}\\
\cmidrule(lr){2-5}\cmidrule(lr){6-7}
system & $\mathrm{TT}$ & $\mathrm{CP}$ & $\mathrm{SEL}$ & $\mathrm{FAI}$
 & \gsym{} & \pone{}\\
\midrule
\NovaS{}  & 0.062$\Downarrow$ & 1.022 & 0.991
          & 1.244$\Downarrow$ & 0.000$\Downarrow$ & 0.000\\
\GemS{}   & 0.538$\Uparrow$ & 1.178$\Downarrow$
          & 0.782$\Uparrow$ & 1.511 & 0.978$\Uparrow$ & 0.489\\
\GptS{}   & 0.561 & 1.289$\Downarrow$ & 0.551$\Uparrow$
          & 1.644$\Downarrow$ & 0.400$\Downarrow$ & 0.200\\
\GptmS{}  & 0.387$\Downarrow$ & 1.156$\Downarrow$ & 0.526$\Uparrow$
          & 1.333$\Downarrow$ & 0.311$\Downarrow$ & 0.089\\
\GrokS{}  & 0.682 & 1.378$\Downarrow$
          & 0.507$\Downarrow$ & 1.489$\Downarrow$ & 0.867$\Uparrow$ & 0.533\\
\midrule
outside range & 3/5 & 4/5 & 4/5 & 4/5 & 5/5 & 0/5\\
\bottomrule
\end{tabular}}
\caption{\textbf{\navworld{} against each system's five-world range.}
$\Uparrow$/$\Downarrow$ mark values above/below the interval the
system spans across the five enterprise worlds
(Table~\ref{tab:main-merged}); unmarked values fall inside it. The left
block is computed by identical code in every world; the right block
substitutes the world's terminal state, so part of any difference is
definitional. The move to navigation lands on how the conversation
goes, not on whether the task is completed.}
\label{tab:bandtable}
\end{table}
\section{Metric--reward associations}
\label{app:corr}
Table~\ref{tab:corr} gives the Spearman rank correlation of every
metric with \pone{} over the 25 enterprise world$\times$system
cells, and over the 20 cells that exclude the disengaged system.
Turn-taking is uncorrelated with the reward ($\rho=-0.06$) and
selectivity negatively correlated ($-0.45$, collapsing to $+0.05$ once
the disengaged system is excluded); conversation progression
($+0.83$), the goal-state factor ($+0.77$) and faithfulness ($+0.58$)
track it, and none of these can be raised by abstaining. The
correlations do not show the abstention metrics are useless among
engaged systems. They show one disengaged system is enough to top a
leaderboard built on them, and a metric a system can win by not working
is broken whether or not anyone exploits it. The exception is the
under-effort share $\pi^{-}$, the strongest process predictor of
success in the suite ($\rho=-0.85$; $-0.73$ engaged-only): a measure
of inactivity that does not reward inactivity, because it is signed
against the task's own reference workload. The MOS predictors go flat
to negative among the engaged four, consistent with
Section~\ref{sec:r1}.
\begin{table}[t]
\centering\footnotesize
\setlength{\tabcolsep}{5pt}
\renewcommand{\arraystretch}{1.1}
\begin{tabular}{@{}lrr@{}}
\toprule
metric & all 25 cells & engaged 20\\
\midrule
\pthree{} & $+0.96$ & $+0.93$\\
$\mathrm{CP}$ & $+0.83$ & $+0.69$\\
\gsym{} & $+0.77$ & $+0.61$\\
$\mathrm{FAI}$ & $+0.58$ & $+0.63$\\
$\widehat{\mathrm{M}}_{\mathrm{U}}$ & $+0.49$ & $+0.02$\\
$\widehat{\mathrm{M}}_{\mathrm{N}}$ & $+0.29$ & $-0.37$\\
$\widehat{\mathrm{M}}_{\mathrm{D}}$ & $-0.04$ & $-0.39$\\
$\mathrm{TT}$ & $-0.06$ & $+0.26$\\
$\mathrm{SEL}$ & $-0.45$ & $+0.05$\\
$\pi^{-}$ & $-0.85$ & $-0.73$\\
\bottomrule
\end{tabular}
\caption{\textbf{Metric--reward associations}: Spearman rank
correlation of each metric with \pone{} over the 25 enterprise
world$\times$system cells (Table~\ref{tab:effort} for $\pi^{-}$), and
over the 20 cells excluding the disengaged system. These are
associations between our own metrics, not against human ratings.}
\label{tab:corr}
\end{table}
\section{Harness dials}
\label{app:harnessdials}
Table~\ref{tab:harness} collects the three configuration dials beside
the system-identity contrast; the closing paragraph of
Appendix~\ref{app:addres} reads them.
\begin{table}[t]
\centering\footnotesize
\begin{tabular}{@{}lcc@{}}
\toprule
dial & levels & $\Delta$ \pone{}\\
\midrule
VAD threshold & 0.2 / 0.5 & personas to full silence\\
step cap & scored / excluded & $+0.086$, $+0.105$\\
simulator LM & 3 models & 0.267 -- 0.578\\
\midrule
\emph{system identity} & 5 systems & 0.00 -- 0.67\\
\bottomrule
\end{tabular}
\caption{Harness dials and effect sizes, with system identity as the
reference contrast; the simulator-LM row is Table~\ref{tab:simablation}; the VAD row is the
persona-audibility effect of Section~\ref{sec:harnesssens} (personas inaudible
at the 0.5 default); the step-cap row is the shift in two systems' pooled
\pone{} when capped runs are excluded rather than scored zero.}
\label{tab:harness}
\end{table}
\section{Duplex phenomenon audit}
\label{app:duplex}
We print this audit rather than leave the claim on the title page
unsupported: a benchmark's honest description includes the gap between
what it elicits and what it scores. The harness elicits the full-duplex
phenomena (user-side barge-in at a fixed cadence, backchannels,
overlapping speech, dead air, and mid-utterance revision) and the
suite scores a subset: $\mathrm{TT}$ scores every floor transfer's
offset and zeroes on unanswered turns, $\mathrm{SEL}$ scores the
response to every injected distractor, and the \textsc{suspension} type
scores dead air across the gap. Elicited but not scored: backchannel
quality and placement, the prosody of overlapping speech, and whether an
agent's mid-utterance revision is fluent or abrupt. The reward never sees
any of these, so a system could be graceless at every overlap and lose
nothing on the agentic pillar, which is why dynamics stays a pillar
beside the reward rather than folded into it, and why no composite is
formed.
\section{Conversation-type excerpts: failure modes}
\label{app:failures}
\section{Enterprise results by conversation type}
\label{app:bytype}
\begin{table*}[t]
\centering\scriptsize
\setlength{\tabcolsep}{2.6pt}
\renewcommand{\arraystretch}{1.06}
\begin{tabular}{@{}ll ccccccccc@{}}
\toprule
metric & system & \axi{1} & \axi{2} & \axi{3} & \axi{4} & \axi{5}
& \axi{6} & \axi{7} & \axi{8} & \axi{9}\\
\midrule
\multirow{5}{*}{\pone{}}
 & \sNova & 0.000 & 0.000 & 0.000 & 0.000 & 0.120 & 0.000 & 0.000 & 0.000 & 0.000\\
 & \sGem  & 0.493 & 0.320 & 0.360 & 0.347 & 0.307 & 0.440 & 0.467 & 0.320 & 0.360\\
 & \sGpt  & 0.640 & 0.533 & 0.573 & 0.440 & 0.507 & 0.413 & 0.440 & 0.333 & 0.440\\
 & \sGptm & 0.320 & 0.107 & 0.160 & 0.267 & 0.227 & 0.187 & 0.200 & 0.107 & 0.293\\
 & \sGrok & 0.680 & 0.493 & 0.453 & 0.360 & 0.413 & 0.467 & 0.667 & 0.440 & 0.360\\
\midrule
\multirow{5}{*}{$\mathrm{ACT}$}
 & \sNova & 0.000 & 0.000 & 0.100 & 0.000 & 0.155 & 0.000 & 0.042 & 0.000 & 0.069\\
 & \sGem  & 0.575 & 0.342 & 0.568 & 0.478 & 0.333 & 0.559 & 0.542 & 0.557 & 0.437\\
 & \sGpt  & 0.841 & 0.597 & 0.671 & 0.750 & 0.582 & 0.625 & 0.567 & 0.629 & 0.651\\
 & \sGptm & 0.475 & 0.192 & 0.492 & 0.435 & 0.297 & 0.294 & 0.386 & 0.250 & 0.492\\
 & \sGrok & 0.720 & 0.569 & 0.613 & 0.561 & 0.389 & 0.662 & 0.667 & 0.606 & 0.500\\
\midrule
\multirow{5}{*}{$\mathrm{NLA}$}
 & \sNova & 0.000 & 0.000 & 0.029 & 0.191 & 0.288 & 0.174 & 0.000 & 0.095 & 0.000\\
 & \sGem  & 0.875 & 0.400 & 0.533 & 0.603 & 0.739 & 0.833 & 0.850 & 0.492 & 0.635\\
 & \sGpt  & 0.811 & 0.684 & 0.685 & 0.774 & 0.828 & 0.857 & 0.667 & 0.623 & 0.762\\
 & \sGptm & 0.515 & 0.444 & 0.210 & 0.596 & 0.542 & 0.409 & 0.375 & 0.391 & 0.450\\
 & \sGrok & 0.825 & 0.778 & 0.627 & 0.612 & 0.712 & 0.750 & 0.850 & 0.556 & 0.731\\
\midrule
\multirow{5}{*}{no-tool}
 & \sNova & 0.560 & 0.293 & 0.613 & 0.547 & 0.693 & 0.493 & 0.520 & 0.467 & 0.560\\
 & \sGem  & 0.013 & 0.000 & 0.013 & 0.000 & 0.080 & 0.093 & 0.000 & 0.027 & 0.027\\
 & \sGpt  & 0.000 & 0.013 & 0.147 & 0.000 & 0.200 & 0.133 & 0.013 & 0.013 & 0.027\\
 & \sGptm & 0.013 & 0.013 & 0.107 & 0.040 & 0.080 & 0.187 & 0.000 & 0.000 & 0.027\\
 & \sGrok & 0.000 & 0.000 & 0.040 & 0.000 & 0.160 & 0.133 & 0.000 & 0.013 & 0.027\\
\midrule
\multirow{5}{*}{\shortstack{WER\textsubscript{cred}\\(\%)}}
 & \sNova & 39.7$^{(25)}$ & 25.0$^{(40)}$ & 36.4$^{(22)}$ & 41.5$^{(29)}$ & 40.4$^{(15)}$ & 32.8$^{(29)}$ & 30.1$^{(28)}$ & 29.0$^{(31)}$ & 33.7$^{(23)}$\\
 & \sGem  & 14.4$^{(59)}$ & 10.5$^{(55)}$ & 17.7$^{(49)}$ & 18.0$^{(65)}$ & 25.2$^{(52)}$ & 12.0$^{(49)}$ & 12.6$^{(55)}$ & 11.7$^{(54)}$ & 16.9$^{(59)}$\\
 & \sGpt  & 7.8$^{(60)}$ & 3.0$^{(55)}$ & 3.5$^{(43)}$ & 17.3$^{(65)}$ & 18.7$^{(49)}$ & 23.0$^{(49)}$ & 7.1$^{(54)}$ & 2.3$^{(54)}$ & 10.5$^{(59)}$\\
 & \sGptm & 11.9$^{(60)}$ & 19.4$^{(54)}$ & 23.2$^{(47)}$ & 19.8$^{(62)}$ & 32.7$^{(49)}$ & 27.5$^{(50)}$ & 22.9$^{(55)}$ & 14.7$^{(55)}$ & 13.1$^{(58)}$\\
 & \sGrok & 14.2$^{(60)}$ & 11.6$^{(55)}$ & 18.4$^{(49)}$ & 16.2$^{(65)}$ & 18.0$^{(49)}$ & 9.2$^{(49)}$ & 11.3$^{(55)}$ & 11.0$^{(54)}$ & 18.9$^{(59)}$\\
\bottomrule
\end{tabular}
\caption{\textbf{Enterprise deep dive by conversation type}, pooled
over the five enterprise worlds (episode-weighted;
WER\textsubscript{cred} conditioned on credential-attempting episodes
with cell $n$ in superscript). $\mathrm{ACT}$ and $\mathrm{NLA}$
inherit the outcome-correlated exclusion of
Table~\ref{tab:convdeep-agents} and are upper bounds. \sNova{}'s
WER\textsubscript{cred} row is conditioned on the minority of its calls
in which it acted at all.}
\label{tab:convdeep-types}
\end{table*}
Table~\ref{tab:convdeep-types} decomposes the enterprise deep dive of
Table~\ref{tab:convdeep-agents} by conversation type.
\section{Run statistics}
\label{app:runstats}
The scored corpus comprises \nconvall{} conversations (3{,}375
enterprise, 450 \navworld{}), totalling 387 hours of simulated
speech. The \navworld{} block is 9 campaigned scenarios $\times$ 5
systems $\times$ 5 runs $\times$ 2 channels, median duration 20.0
minutes; 215 of its 450 conversations end at the step cap and none at the
wall-clock cap (Appendix~\ref{app:navcensor}), and its realistic channel
carries the headline cells (225 conversations). In the enterprise
worlds, 302 conversations end at the wall-clock timeout; every capped or
timed-out run scores reward 0 and remains in the corpus
(Section~\ref{sec:harness}).
\section{Results in pictures}
\label{app:pictures}
\begin{figure*}[t]
\centering
\includegraphics[width=\textwidth]{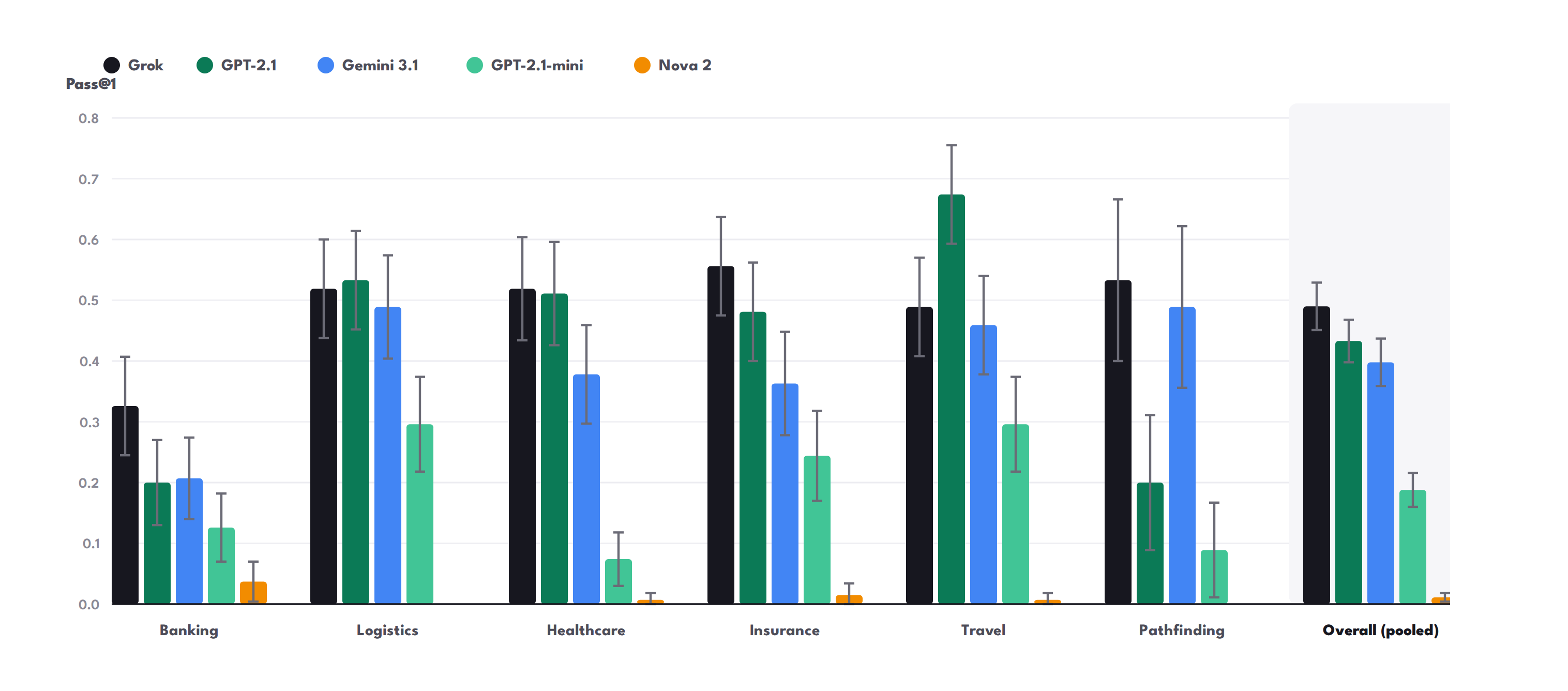}
\caption{\pone{} by world and pooled overall; whiskers are 95\%
bootstrap intervals.}
\label{fig:worldbars}
\end{figure*}

\begin{figure*}[t]
\centering
\includegraphics[width=\textwidth]{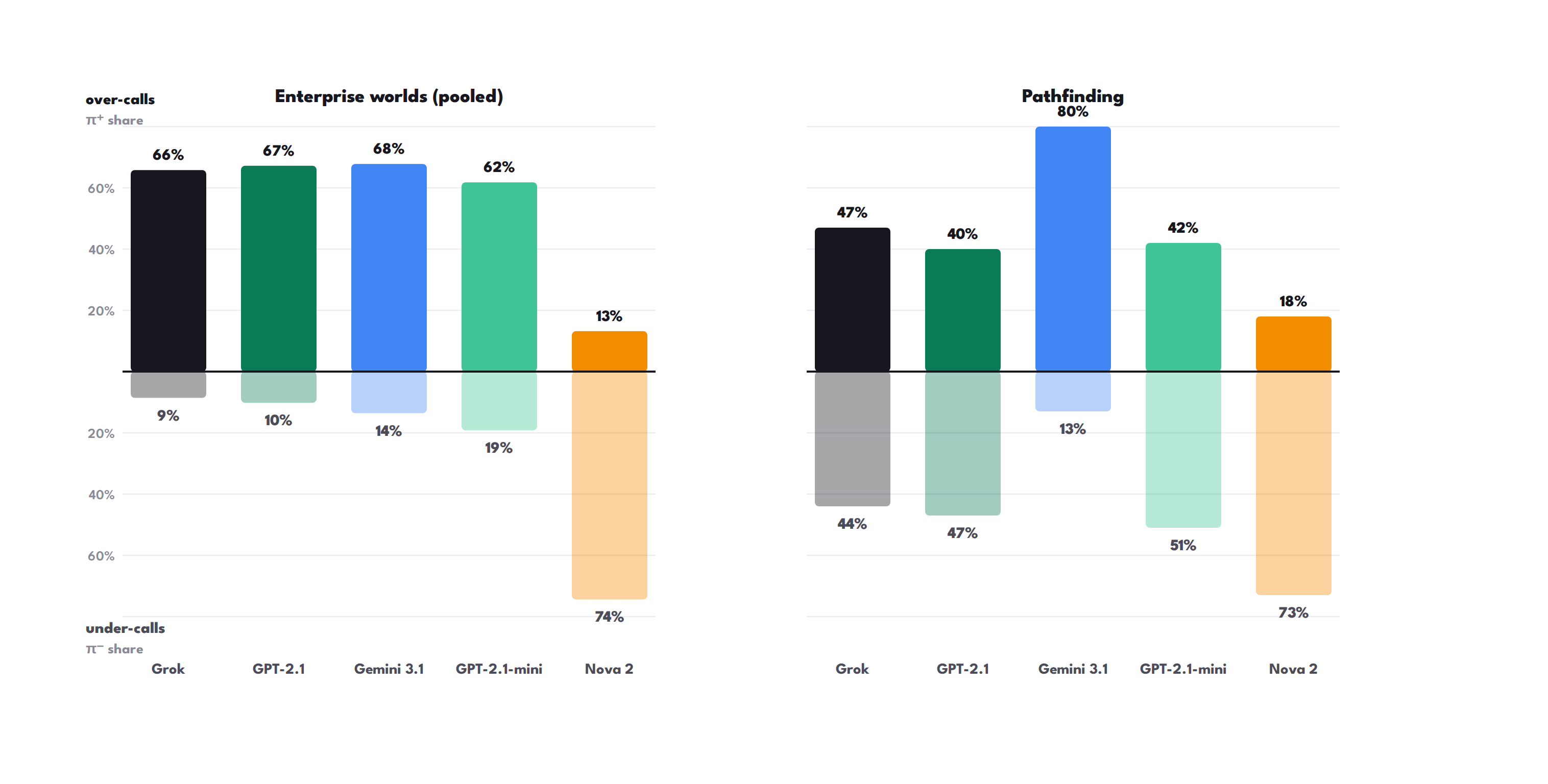}
\caption{Share of conversations issuing more (solid, up) or fewer
(faded, down) tool calls than the reference workload.}
\label{fig:effortbars}
\end{figure*}

\begin{figure}[t]
\centering
\includegraphics[width=\columnwidth]{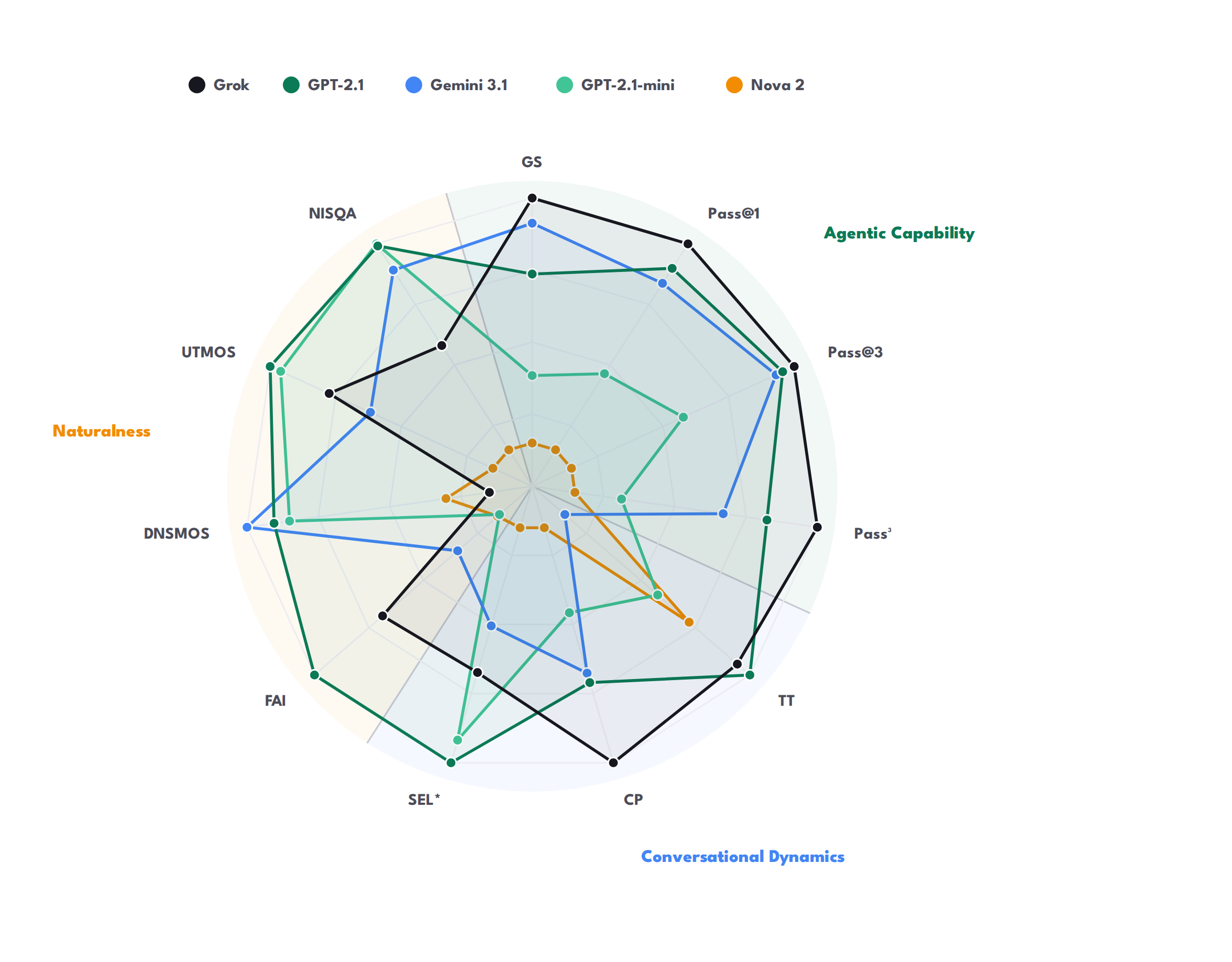}
\caption{All eleven metrics of Table~\ref{tab:main-merged}, pooled
over worlds and min--max scaled per axis; larger is better (SEL
inverted). Sectors mark the three pillars.}
\label{fig:radar11}
\end{figure}

\begin{figure}[t]
\centering
\includegraphics[width=\columnwidth]{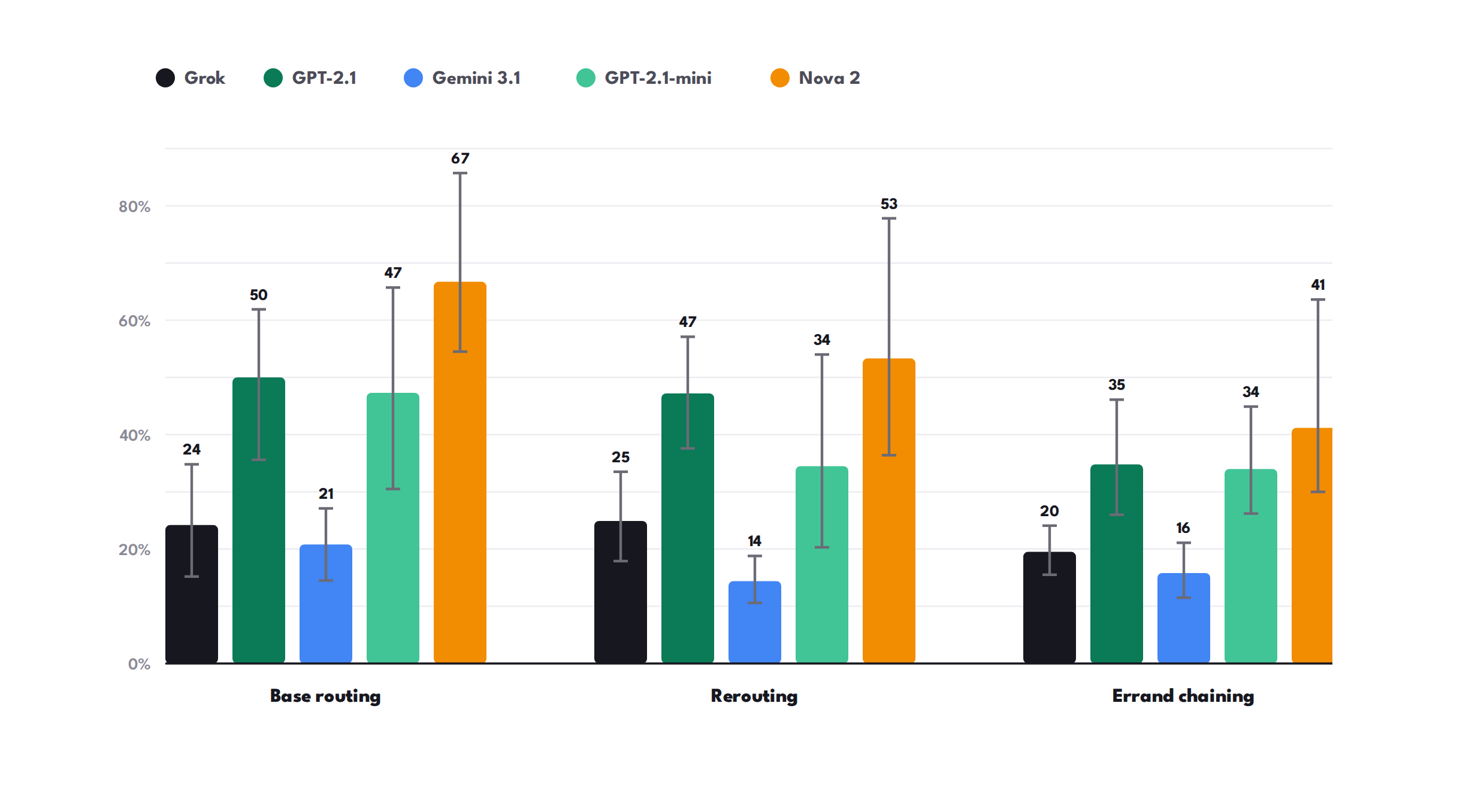}
\caption{Exploration share in \navworld{} by system and conversation
type (realistic channel); whiskers are 95\% bootstrap intervals.}
\label{fig:exploretypes}
\end{figure}

Figures~\ref{fig:worldbars}--\ref{fig:exploretypes} redraw the
headline numbers of Section~\ref{sec:results} for visual comparison.
\section{License of Artifacts}
\label{app:licenses}
All artifacts are used for non-commercial academic research, consistent
with their terms. The five evaluated systems are proprietary services
accessed through their providers' APIs under the corresponding terms of
service: \Nova{} via Amazon Bedrock, \Gem{} via the Google Gemini API,
\Gpt{} and \Gptm{} via the OpenAI API, and \Grok{} via the xAI API.
The harness-side models are likewise proprietary services: the user
simulator \texttt{gpt-5.6-luna} via Azure, and the decision model
\texttt{claude-haiku-4.5} and the judge models (\texttt{gpt-5.2},
\texttt{claude-opus-4.6}, \texttt{grok-4.3}) via OpenRouter under each
provider's terms; caller and walker speech is synthesised through the
ElevenLabs API under its terms. No provider model weights are
redistributed. Among open-source evaluation components, the $\tau$-bench
and $\tau^2$-bench harnesses our orchestrator descends from are
MIT-licensed, as is the DNSMOS code in the DNS-Challenge repository and
the UTMOS (UTMOS22) system; NISQA is MIT-licensed code with pretrained
weights under CC BY-NC-SA 4.0, which our non-commercial use respects;
the turn-taking metric is reimplemented from EVA-Bench's published
specification \citep{evabench2026}. Telephony coding follows the public
ITU-T G.711 standard. All scenarios, records, policies, personas and
identifiers in \bench{} are authored by us and contain no third-party
data, and the realistic channel's background-noise recording inventory
is distributed with the release. We release the scenario corpus,
annotations and evaluation code for research use, under CC BY 4.0 for
data and MIT for code.
\section{Use of AI Assistants}
\label{app:ai}
AI assistants were used in the writing and coding process, and not as
part of the research itself. In writing, assistance was limited to
grammar and spell checking. In coding, an AI-assisted IDE (Claude Code)
was used during implementation. Task design, scenario authoring,
experiments, analyses and conclusions are the authors' own.
\end{document}